\documentclass{aa}
\usepackage{graphicx,times}

\begin{document}

\title{Spatially resolved X-ray spectroscopy of cooling clusters of galaxies}

\author{ J.S. Kaastra \inst{1}
         \and
         T. Tamura \inst{1}
         \and
         J.R. Peterson \inst{2}
         \and
         J.A.M. Bleeker \inst{1}
         \and
         C. Ferrigno \inst{1}
         \and
         S.M. Kahn \inst{2}
         \and
         F.B.S. Paerels \inst{2}
	\and 
	 R. Piffaretti \inst{3,4}	
	\and
         G. Branduardi-Raymont \inst{5} 
         \and
         H. B\"ohringer \inst{6}
         }
  
\offprints{J.S. Kaastra}
\mail{J.Kaastra@sron.nl}

\institute{ SRON National Institute for Space Research
              Sorbonnelaan 2, 3584 CA Utrecht, The Nether\-lands 
              \and
              Department of Physics, Columbia University,
              550 West 120th Street, New York, 
              NY 10027, USA
	      \and
	      Paul Scherrer Institute, Laboratory for Astrophysics, 
	      CH-5232 Villigen, Switzerland
              \and 
              Institute of Theoretical Physics, University of Z\"{u}rich, 
              Winterthurerstrasse, 190, CH-8057 Z\"{u}rich, Switzerland 
              \and
	      Mullard Space Science Laboratory,
              University College London, Holmbury St. Mary, Dorking,
              Surrey, RH5 6NT, UK 
              \and
              Max-Planck-Institut f\"ur extraterrestrische Physik, 
              Giessenbachstrasse, 85748 Garching, Germany
              }

\date{Received  / Accepted  }

\abstract{
We present spatially resolved X-ray spectra taken with the EPIC cameras of
XMM-Newton of a sample of 17 cooling clusters and three non-cooling clusters for
comparison.  The deprojected spectra are analyzed with a multi-temperature
model, independent of any a priori assumptions about the physics behind the
cooling and heating mechanisms.  All cooling clusters show a central decrement
of the average temperature, most of them of a factor of $\sim 2$.  Three
clusters (S\'ersic~159$-$3, MKW~3s and Hydra~A) only show a weak temperature
decrement, while two others (A~399 and A~2052) have a very strong temperature
decrement.  All cooling clusters show a weak pressure gradient in the core.
More important, at each radius within the cooling region the gas is not
isothermal.  The differential emission measure distribution shows a strong peak
near the maximum (ambient) temperature, with a steep decline towards lower
temperatures, approximately proportional to $T^{3}$, or alternatively a cut-off
at about a quarter to half of the maximum temperature.  In general, we find a
poor correlation between radio flux of the central galaxy and the temperature
decrement of the cooling flow.  This is interpreted as evidence that except for
a few cases (like the Hydra A cluster) heating by a central AGN is not the most
common cause of weak cooling flows.  We investigate the role of heat conduction
by electrons and find that the theoretically predicted conductivity rates are
not high enough to balance radiation losses.  The differential emission measure
distribution has remarkable similarities with the predictions from coronal
magnetic loop models.  Also the physical processes involved (radiative cooling,
thermal conduction along the loops, gravity) are similar for clusters loops and
coronal loops.  If coronal loop models apply to clusters, we find that a few
hundred loops per scale height should be present.  The typical loop sizes
deduced from the observed emission measure distribution are consistent with the
characteristic magnetic field sizes deduced from Faraday rotation measurements.

}

\maketitle

\section{Introduction}

The visible mass in clusters of galaxies is dominated by hot diffuse gas.  Due
to the high temperature of the gas it is visible in the X-ray band.  In the
central part of the cluster the gas density is high.  There the radiative
cooling time is shorter than the age of the cluster and the age of the universe.
The temperature decreases due to radiative cooling.  The corresponding pressure
decrease then causes a net inflow towards the center of the cluster.  A review
of this so-called cooling flow process is given by Fabian (\cite{fabian94}).
The associated mass deposition rate $\dot{M}$ may reach several thousands
M$_{\sun}$ per year in the strongest cooling flows.  In general, the time the
gas needs to cool down to very low temperatures is smaller than the flow time
towards the center of the cluster.  Thus, gas drops out locally from the cooling
flow.

In the isobaric cooling flow model (Johnstone et al.  \cite{johnstone92}) gas
cools in pressure equilibrium with its surroundings.  The amount of gas at each
temperature is controlled by how fast it can cool, i.e.  the radiative cooling
rate through -- mainly -- X-ray radiation.  The mass deposition rate is not the
same at all radii but scales linearly with the radius, $\dot{M}\sim r$.

The first high-resolution X-ray spectra of clusters of galaxies taken with the
Reflection Grating Spectrometers (RGS) of XMM-Newton showed indeed the presence
of cooler gas in the cores of several clusters.  However, the amount of cool gas
at lower temperatures was much smaller than predicted by the isobaric cooling
flow model (S\'ersic~159$-$3, Kaastra et al.  \cite{kaastra01}; A~1835,
Peterson et al.  \cite{peterson01}; A 1795, Tamura et al.  \cite{tamura01a}).
The second paper also contains a list of possible explanations for this
phenomenon.

This discovery has triggered a series of papers offering a broad range of
explanations for the apparent failure of the isobaric cooling flow model.
Worked-out examples include metallicity inhomogeneities (Fabian et al.
\cite{fabian01}), buoyantly rising radio bubbles transporting cool gas outwards
(Churazov et al.  \cite{churazov01}), halo-in-halo structure (Makishima et al.
\cite{makishima01}), turbulent mixing due to rising and falling hot gas bubbles
(Quilis et al.  \cite{quilis01}), heating by AGN activity (B\"ohringer et al.
\cite{boehringer02}), contamination due to non-thermal X-ray emission (McCarthy
et al.  \cite{mccarthy02}), heating due to dead radio galaxies (Reynolds et al.
\cite{reynolds02}), rapid cooling due to mixing with cold gas (Fabian et al.
\cite{fabian02}), and heat conduction by electrons (Voigt et al.
\cite{voigt02}).

The lack of relatively cool gas has been confirmed by the RGS spectra of other
clusters, like A~496 (Tamura et al.  \cite{tamura01b}) and Virgo (Sakelliou et
al.  \cite{sakelliou02}).  The lack of this cool gas is deduced from both RGS
data as well as from XMM-Newton EPIC data.  These EPIC data have also been
reported in the RGS papers mentioned above.  In addition, Molendi \& Pizzolato
(\cite{molendi01}) found evidence for an apparent low temperature cut-off in the
cooling gas at about 1--2~keV in the EPIC spectra of A~1835, A~1795 and Virgo.
Matsushita et al.  (\cite{matsushita02}) report that the EPIC data of Virgo
show that the cool gas has a single temperature at each radius.

Also Chandra observations have confirmed the lack of cool gas in clusters.  In
Hydra~A (David et al.  \cite{david01}) the gas has essentially a single
temperature at each radius between 30--200~kpc, with a steadily outwards
temperature increase.  Only in the innermost 30~kpc there is evidence for
multiphase gas, however with a ten times smaller mass deposition rate as derived
from the morphologically derived mass accretion rate at 30~kpc.

The superb energy resolution of the RGS allowed a detailed investigation of the
temperature structure of the core.  Although the RGS has spatial resolution in
the cross dispersion direction of the gratings, for most of the available data
the statistics are not good enough to map the cool gas in the cross dispersion
direction.  In order to be able to distinguish between the various theoretical
models that have been proposed, it is important to measure the temperature
structure for each radius.  As shown above, for some clusters a single-phase
cooling gas distribution at each radius has been measured (Hydra~A, David et al.
(\cite{david01}); Virgo, Matsushita et al.  (\cite{matsushita02}); A~2029,
Lewis et al.  (\cite{lewis02}).

In this paper we study a large sample of clusters and use the spatially resolved
XMM-Newton/EPIC spectra to study the temperature structure at each radius.
These measurements allow us to distinguish between isothermal, isobaric cooling
or cut-off cooling flow models.

The layout of this paper is as follows.  In Sect.~\ref{sect:data} we give a
description of our data analysis procedures.  This description is extensive
since it also forms the basis of a set of papers on other properties of these
clusters that are in preparation (e.g., soft excess emission (Kaastra et al.
\cite{kaastra03}); abundances (Tamura et al., in preparation).  In
Sect.~\ref{sect:results} we describe the results of our spectral modeling using
several models, and in Sect.~\ref{sect:discussion} we compare our results to a
range of physical models for the cooling gas.  Throughout this paper we use $H_0
= 50$~km\,s$^{-1}$\,Mpc$^{-1}$ and $q_0=0.5$.

\section{Data analysis\label{sect:data}}

\subsection{Cluster sample}

Our sample consists of 17 clusters with cooling flows, taken from the Guaranteed
Time program of the RGS consortium and other available data sets.  These cooling
flow clusters were selected mainly based upon their suitability for
spectroscopic studies with the Reflection Grating Spectrometers (RGS) of
XMM-Newton.  This selection criterion favors relatively medium distant, compact
and cool clusters.  Nearby clusters in general have a large angular size,
resulting in a degradation of the spectral resolving power of the RGS.  In very
hot clusters the line emission is weak with respect to the Bremsstrahlung
continuum.  In addition we added three non cooling clusters as a control case
for clusters without cooling gas:  the Coma cluster and A~3266, as well as the
merging cluster A~754.  The cluster sample is listed in Table~\ref{tab:cluspar}.
We list in this and all subsequent tables the clusters in increasing hot gas
temperature order.

\begin{table}[!ht]
\caption{Basic cluster parameters}
\label{tab:cluspar}
\centerline{
\begin{tabular}{|lrrrrr|}
\hline
Cluster               & $z$\,$^{\mathrm{a}}$&  scale\,$^{\mathrm{b}}$ & $kT$\,$^{\mathrm{c}}$
     & $F_{\mathrm X}$\,$^{\mathrm{d}}$ & $N_{\mathrm H}$\,$^{\mathrm{e}}$\\
\hline
NGC 533               & 0.0175& 30& 1.3&   6& 3.00 \\
Virgo                 & 0.0027&4.7& 2.4& 821& 1.80 \\
A 262                 & 0.0155& 26& 2.4&  49& 8.94 \\
A 1837                & 0.0707&109& 2.4&   7& 4.38 \\              
S\'ersic~159$-$3      & 0.0572& 90& 2.5&  24& 1.79 \\
MKW 9                 & 0.0402& 65& 2.7& 2.3& 4.18 \\
2A 0335+096           & 0.0344& 57& 3.0&  81&28.71 \\
MKW 3s                & 0.0455& 73& 3.0&  30& 2.89 \\
A 2052                & 0.0356& 58& 3.1&  47& 2.91 \\
A 4059                & 0.0466& 75& 3.5&  31& 1.06 \\
Hydra A  (A 780)      & 0.0550& 87& 3.8&  48& 4.80 \\
A 496                 & 0.0322& 53& 3.9&  75& 6.44 \\
A 3112                & 0.0756&116& 4.1&  36& 2.61 \\
A 1795                & 0.0639&100& 5.3&  68& 1.01 \\
A 399                 & 0.0706&109& 5.8&  29&10.90 \\
A 3266                & 0.0614& 96& 6.2&  49& 1.60 \\
Perseus (A 426)       & 0.0179& 30& 6.3& 926&14.90 \\
Coma (A 1656)         & 0.0240& 40& 8.1& 319& 0.89 \\
A 754                 & 0.0561& 89& 9.1&  64& 5.67 \\
A 1835                & 0.2541&298&14.8&  15& 2.32 \\
\hline\noalign{\smallskip}
\end{tabular}
}
\begin{list}{}{}
\item[$^{\mathrm{a}}$] Redshift, to be used for the distance estimate.
\item[$^{\mathrm{b}}$] Angular scale, in kpc/arcmin.
\item[$^{\mathrm{c}}$] Temperature, in keV.
\item[$^{\mathrm{d}}$] Unabsorbed 0.1--2.4~keV X-ray flux, in $10^{-15}$~W\,m$^{-2}$.
\item[$^{\mathrm{e}}$] Galactic column density, in $10^{24}$~m$^{-2}$.
\end{list}
\end{table}

Temperatures in Table~\ref{tab:cluspar} were taken from Ebeling et al.
(\cite{ebeling98}) for NGC~533 and A~1835, from Finoguenov et al.
(\cite{finoguenov01} for MKW~9 and from David et al.  (\cite{david93}) for all
other clusters.  Unabsorbed X-ray fluxes (0.1--2.4~keV) were taken from Ebeling
et al.  (\cite{ebeling96}) for A\,426, A\,496, A\,780, A\,1837, and A\,4059;
from Cruddace et al.  (\cite{cruddace02}) for S\'ersic~159$-$3; from Kriss et
al.  (\cite{kriss83}) for MKW~9; and from Ebeling et al.  (\cite{ebeling98}) for
all others.  Note that the data presented for A~189 by Ebeling et al.
(\cite{ebeling98}) correspond to NGC~533 (as analyzed in this paper), not to
A~189.

\subsection{Data extraction}

\begin{table*}[!ht]
\caption{Observation log}
\label{tab:obslog}
\centerline{
\begin{tabular}{|l|r|rrr|rr|rrr|ll|rrr|}
\hline
\multicolumn{1}{|l|}{Cluster} & \multicolumn{1}{|l|}{XMM} &
   \multicolumn{3}{|c|}{Exposure time (ks)$^{\mathrm{a}}$} &
    \multicolumn{2}{|c|}{$H_t$$^{\mathrm{b}}$} &
    \multicolumn{3}{|c|}{Average $H$$^{\mathrm{c}}$ } &
    \multicolumn{1}{|c}{mode$^{\mathrm{d}}$} & 
    \multicolumn{1}{c|}{filter$^{\mathrm{e}}$} \\
name                  & orbit & MOS1 & MOS2 & pn & MOS & pn & MOS1 & MOS2 & pn & pn & \\
\hline
NGC 533               & 195 & 38 & 38 & 31 & 35 & 50 & 21.8 & 22.3 & 28.9 & FF & thin \\
Virgo                 &  97 & 34 & 33 & 25 & 35 & 50 & 25.6 & 26.2 & 36.8 & FF & thin \\
A 262                 & 203 & 24 & 24 & 15 & 35 & 50 & 25.1 & 24.6 & 33.0 & FF & thin \\
A 1837                & 200 & 48 & 48 & 40 & 35 & 50 & 21.5 & 22.3 & 29.7 &EFF & thin \\	  
S\'ersic~159$-$3      &  77 & 31 & 32 & 28 & 35 & 50 & 25.5 & 24.7 & 32.5 & FF & thin \\
MKW 9                 & 311 & 22 & 23 & 13 & 35 & 50 & 27.9 & 28.4 & 31.1 &EFF & thin \\
2A 0335+096           & 215 & 11 & 11 &  5 & 50 & 56 & 28.4 & 29.5 & 27.2 & FF & thin \\
MKW 3s                & 129 & 37 & 37 & 32 & 35 & 50 & 22.1 & 22.5 & 31.2 & FF & thin \\
A 2052                & 128 & 30 & 30 & 23 & 35 & 50 & 21.4 & 21.7 & 29.6 &EFF & thin \\
A 4059                & 176 & 13 & 13 &  6 & 50 & 56 & 33.2 & 32.3 & 34.1 & FF & thin \\
                      & 176 & 21 & 20 &  - & 40 &  - & 28.2 & 28.8 &    - & FF & thin \\
Hydra A  (A 780)      & 183 & 17 & 17 &  3 & 50 & 56 & 32.8 & 32.4 & 31.6 & FF & thin \\
A 496                 & 211 &  9 & 10 &  5 & 35 & 56 & 27.3 & 25.6 & 38.9 &EFF & thin \\
A 3112                & 191 & 22 & 22 & 16 & 35 & 50 & 24.7 & 24.3 & 34.4 &EFF & med  \\
A 1795                & 100 & 29 & 30 & 22 & 35 & 52 & 25.9 & 25.8 & 38.8 & FF & thin \\
A 399                 & 127 & 13 & 13 &  6 & 35 & 50 & 22.2 & 23.9 & 31.3 &EFF & thin \\
A 3266                & 153 & 24 & 24 & 19 & 35 & 50 & 24.0 & 21.7 & 28.9 &EFF & med  \\
Perseus (A 426)       & 210 & 43 & 46 & 33 & 80 & 120& 62.5 & 64.9 & 82.5 & FF & thin \\ 
Coma (A 1656)         &  86 & 15 & 16 & 13 & 35 & 50 & 24.7 & 24.9 & 42.9 & FF & med  \\
A 754                 & 262 & 14 & 14 & 11 & 35 & 50 & 24.2 & 24.6 & 34.4 &EFF & med  \\
A 1835                & 101 &  - & 26 & 25 & 35 & 50 &    - & 24.2 & 33.4 & FF & thin \\
\hline
Background            &  -  & 410& 379& 324& 60 & 50 & 23.8 & 23.1 & 30.8 & FF & thin \\ 
\hline\noalign{\smallskip}
\end{tabular}
}
\begin{list}{}{}
\item[$^{\mathrm{a}}$] Net exposure time, after rejection of soft proton flares
\item[$^{\mathrm{b}}$] Background rejection threshold for soft proton
                       flares, in counts per 260~s between 10--12~keV
\item[$^{\mathrm{c}}$] Average background after rejection of
                       soft proton flares, in counts per 260~s
                       between 10--12~keV
\item[$^{\mathrm{d}}$] mode pn: extended full frame (EFF) of full frame (FF)
\item[$^{\mathrm{e}}$] filter used for MOS and pn
\item[$^{\mathrm{f}}$] Number of rejected point sources
\end{list}
\end{table*}

Details about the observations and data processing are given in
Table~\ref{tab:obslog}.  For A~1835 we discarded the MOS1 data (taken in a large
window mode).  The data were extracted using the standard SAS software,
equivalent to version 5.3.  Only events with {\sl pattern}$<$13 (MOS) or {\sl
pattern}$=0$ (pn) were accepted; in addition, we selected {\sl flag}$=0$.

\subsection{Background subtraction\label{sect:back}}

The background spectrum of the XMM-Newton EPIC camera's is described in detail
by Lumb et al.  (\cite{lumb02}).  The main components are the cosmic X-ray
background, dominating the background at low energies, and a time-variable
particle component, dominating the high energy background.
Background subtraction is always done using the same detector regions for
source and background.

\subsubsection{Time-variable particle background}

The time-variable particle component is caused by clouds of soft protons.  Apart
from long quiescent intervals, there are sporadic episodes where this background
component is enhanced by orders of magnitude, as well as time intervals with
only a low level of flaring.

The contribution of X-ray photons to the full field 10--12~keV band count rate
is almost negligible as compared with the background.  This energy band is
dominated by low energy protons.  Therefore it can be used as a monitor for the
time-variable background component. 

Thus we divided the observation into time intervals of 260~s, and for each time
interval $H$, the number of counts in the 10--12~keV band was determined.
During quiescent periods, the average value for $H$ is typically 23--24 for MOS
and 31 for pn.  We excluded time intervals with high particle background as
measured by $H>H_t$ where for most observations we took the threshold $H_t=35$
for MOS and $H_t=50$ for pn.  This threshold is about 2.5 and 3.5$\sigma$ above
the average quiescent level, respectively.  Therefore the fraction of time bins
rejected due to Poissonian fluctuations in $H$ is very small.  For some clusters
we had to take slightly higher thresholds in order to retain a minimum of
exposure time.  The thresholds used as well as the background after filtering
are listed also in Table~\ref{tab:obslog}.

The above procedure is sufficient to remove the large flares, simply by
discarding the time intervals with high 10--12~keV count rates.  However after
removing these intense large proton events the average background level
still varies slightly.  The r.m.s.  variation from cluster to cluster is
$\sim$10--20~\%.  This is due to the fact that for some observations the count
rate is slightly enhanced but smaller than the cutoff threshold.  This holds in
particular for the clusters with a higher threshold but also to some extent for
the others.  These remaining weak flares cannot be removed by time selection.
For them, however, the increase in background flux in the 0.2--10~keV raw
spectrum is to first order proportional to the increase in 10--12~keV count
rate, as we illustrate below.

\begin{figure}
\resizebox{\hsize}{!}{\includegraphics[angle=-90]{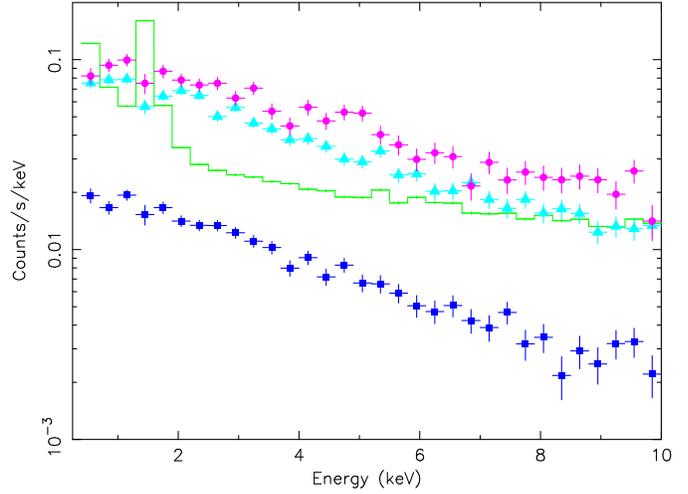}}
\caption{Background spectra for MOS1. Solid histogram:
background spectrum for $H$ between 19--28; squares, triangles and stars:
difference of background spectra for $H$ between 29--38, 39--48 and 49--58,
respectively, with the 19--28 spectrum.}
\label{fig:bgplot1}
\end{figure}

We show in Fig.~\ref{fig:bgplot1} the full-field MOS1 spectrum of our background
field (discussed in the next section) for all 977 bins with $H$ between 19--28,
close to the average value of 23.82.  The soft excess due to the cosmic X-ray
background as well as the instrumental fluorescence lines near 1.5~keV are
clearly seen.  We also show the difference spectra as compared to this "average"
spectrum for $H$ ranges of 29--38 (average value 31.84, for 249 time bins);
39--48 (average value 42.35, for 37 time bins); and 49--58 (average value 53.17,
for 12 time bins).  These difference spectra are smoother than the average
background spectrum, and they are approximately exponential
$\sim\exp({-E/E_c})$, with $E_c$ values of typically 4.2, 4.5 and 5.2 from low
to high background values, respectively.  Thus, as a first approximation, for
small enhancements of the background level, the shape of the background
enhancement as a function of energy is more or less fixed while its amplitude
scales with the difference of $H$ with the average value of $H$.

We utilise this behaviour to make a first order correction to the estimated
background spectrum.  The usual procedure would be to subtract the background
spectrum using a blank field observation screened in the same way as the cluster
data.  Here in order to correct for the weak variability we divide the cluster
and blank field observation into subsets with the same 10--12~keV count rate
(same $H$). 
The subtracted background $B(E)$ was then determined as
\begin{equation}
B(E) = \sum\limits_{H}^{} B_{H}(E) [f_c(H)/f_b(H)][t_c/ t_b],
\end{equation}
where $B_{H}(E)$ is the histogram of background events at a given 10--12~keV
count rate level $H$ in the blank field, $f_c(H)$ and $f_b(H)$ are the frequency
distributions of $H$ for the cluster and blank field, respectively, and $t_c$
and $t_b$ are the exposure times of the cluster and blank fields.
This ensures correct particle background subtraction, under the
assumption that for low proton count rates the shape of the proton spectrum does
not vary when its flux increases by a small amount.  However in order to account
for any possible remaining particle background subtraction problems, we have
included a systematic uncertainty of at least 10~\% of the total background in
all our fits. 

The deep field used in our background subtraction procedure (Lumb et al.
(\cite{lumb02}) consisted of the background event file provided by the
XMM-Newton Science Operations Center, containing a total of 320--410~ks exposure
time on 8 deep fields (the exposure time differs slightly for the different
instruments).  This background event file was filtered the same way as the
source file, except that we took the threshold $H_t$ slightly higher in order
to have some estimate of the background for mildly enhanced soft proton levels
(see Table~\ref{tab:obslog}).

As stated above, the contribution of X-ray photons to the 10--12~keV band count
rate is almost negligible.  There are three possible exceptions to this:  Coma,
Virgo and Perseus.

For Coma, the second brightest cluster in our sample and a very hot cluster, the
full-field 10--12~keV cluster emission reaches a level of 10~\% of the quiescent
background count rate in the pn camera only.  Owing to the lower sensitivity of
the MOS cameras at high energies, the relative cluster contamination in the
10--12~keV band is at most 5~\% there.  Thus, we expect that our method
overestimates the time-variable particle contribution in Coma by 5--10~\%.  This
does not affect our science results for Coma, however, since (i) the difference
is within the systematic background uncertainty used by us; (ii) at lower
energies the background is dominated by the constant cosmic X-ray background,
and not by the soft proton component; (iii) except for the outermost annulus,
the X-ray flux of Coma is a factor of 5--100 times brighter than the subtracted
background for all energies below $\sim$6~keV; (iv) the spectral fit is
dominated by the higher signal-to-noise part of the spectrum below $\sim$6~keV.

Although Virgo is brighter than Coma (Table~\ref{tab:cluspar}), its temperature
is so much lower that the amount of 10--12~keV photons can be neglected safely
with respect to the soft proton background.

Only Perseus has really an enhanced 10--12~keV count rate, as is evident from
the large average values of $H$ as listed in Table~\ref{tab:obslog}.  Here we
subtracted first a constant offset from the 10--12~keV light curves, to be
consistent with the average background level in the deep fields.  Note that
Churazov et al.  (\cite{churazov03}) suggest that for this observation there is
a steady enhanced soft proton background.  However, Perseus is so bright that
in none of our spectra the subtracted background is important.  Moreover, our
results for the cooling region are most sensitive to Fe-L emission, and this is
the spectral range where the background is relatively very low.

\subsubsection{Cosmic X-ray background}

While at high energies the particle contribution dominates the background
spectrum, at low energies the cosmic X-ray background yields the largest
contribution to the measured background.  The cosmic X-ray background varies
from position to position on the sky.  This is in particular important at low
energies.  To get a typical estimate for the sky variation, we took the PSPC
count rates for 378 regions at high Galactic latitude as studied by Snowden et
al.  (\cite{snowden00}) and determined the population variation for this
component in the R1 (low energy) band; the r.m.s.  variation is typically
35$\pm$3~\%.  In principle, the background estimate for any location on the sky
can be improved by taking the measured PSPC count rates into account.  However,
this requires detailed spectral modeling of the soft X-ray background for each
position on the sky that is being studied, which can be rather uncertain due to
the poor spectral resolution of the PSPC.  Here we have taken a conservative
approach by using the average background as contained in the standard EPIC
background files, but we included a systematic background error of 35~\% of the
total background below 0.5~keV, 25~\% between 0.5--0.7~keV, 15~\% between
0.7--2~keV, and 10~\% above 10~keV.  This last 10~\% also includes the
systematic uncertainty in the subtracted time-variable particle background
component.

Note that the above 35~\% systematic background uncertainty at low energies is a
r.m.s.  estimate.  In individual cases there may be a larger deviation from the
average high latitude X-ray background.

The differences between the spectral response for pn data taken with the full
frame mode (as in our deep fields) and the extended full frame mode (as in some
of our clusters) are negligible compared to the systematic uncertainties on the
background and effective area that we use in this paper.  Of more importance is
the difference between data taken with the medium filter (Coma and a few other
clusters) and the thin filter (most of our clusters and our deep field).  For a
power law photon spectrum with photon index 2, the medium filter produces about
10~\% less counts at 0.2~keV, 6~\% less at 0.3~keV, and less than 2~\% above
0.4~keV, as compared to the thin filter.  These differences are all well within
our adopted systematic uncertainties.

Finally, in all fields the brightest X-ray point sources were removed.  This was
done by making an image, sampled with a pixel size of 4\arcsec.  At each pixel,
the net flux within a square of size 5 pixels (20\arcsec) was determined and
compared with the local background as measured in the surrounding square of
17$\times$17 pixels (68$\times$68\arcsec).  All sources stronger than 4$\sigma$
(MOS) or 4$\sqrt{2}\sigma$ (pn) were marked.  Events within a square of
32$\times$32\arcsec\ around the marked sources were discarded.  This procedure
was only done outside a radius of 2\arcmin\ from the cluster center.  Typically,
a median of 8 point sources (for MOS1 and MOS2) or 15 (for pn) was discarded
this way.  In general, we detect more point sources in the high signal to noise
data.  For those clusters point source rejection is indeed relatively important.
Our sensitivity is not seriously affected by this selection, since even for
Virgo the discarded detector area is only 5~\% of the total detector area.

\subsection{Radial profiles and psf effects\label{sect:radprof}}

\begin{figure}
\resizebox{\hsize}{!}{\includegraphics[angle=-90]{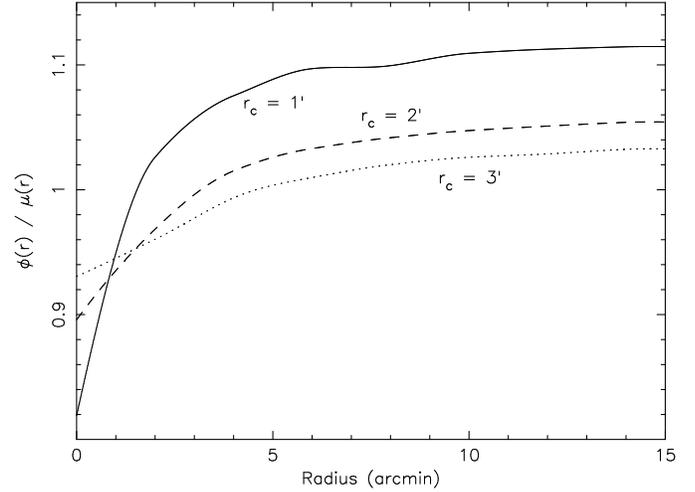}}
\caption{Ratio of psf-convolved cluster profile $\phi(r)$ over
true cluster profile $\mu(r)$ for MOS1 and a photon energy of 1.5~keV.
For $\mu(r)$ we have taken a $\beta$ profile with $\beta=2/3$ and
core radii as indicated.}
\label{fig:phimu}
\end{figure}

The  psf of the
X-ray telescopes of XMM-Newton has a rather narrow core (FWHM of order
6\arcsec), but extended wings.  In fact, the shape of the psf closely resembles
the shape of a typical cluster radial intensity distribution  (see
for instance Ghizzardi \cite{ghizzardi01}).  Thus, spectra
accumulated in a given region of the detector are contaminated by flux from
other regions of the cluster, which may have different spectral shapes.  In this
section we make quantitative estimates of this effect.  For simplicity we assume
here spherical symmetry for both the source radial profile $\mu(r)$ and the
instrumental psf.  We determine how much the observed radial intensity profile
$\phi(r)$ differs from the true radial profile $\mu(r)$.  We denote the
encircled energy fraction of the psf by $F(r)$, with by definition $F(0)=0$ and
$F(\infty)=1$.  The encircled energy $F(r)$ of X-ray telescope 1 (MOS1) can be
described quite accurately with a King profile:
\begin{equation}
\label{eqn:xrt1}
F(r) = 1 - \left[ 1 + (r/a)^2 \right]^{-b},
\end{equation}
where for photon energies of 1.5~keV $a=4.8$\arcsec\ and $b=0.45$, and for 6~keV
photons $a=3.6$\arcsec\ and $b=0.40$.  Unfortunately the distribution
(\ref{eqn:xrt1}) has no finite variance.  For example, if we cut-off all photons
outside a radius of 7.5, 15, 30 and 60\arcmin, the equivalent 1-dimensional
r.m.s.  width $\sigma$ equals 0.61, 0.90, 1.32 and 1.93\arcmin, respectively
(for an energy of 1.5~keV).  This is much larger than the core of the psf, and
comparable to the typical core radius of the clusters we are interested in.
Thus, the psf effects are potentially large.

To illustrate this, we have calculated using a Monte-Carlo simulation with
$10^7$ photons the ratio $\phi(r)/\mu(r)$ for MOS1 (Fig.~\ref{fig:phimu}).  For
$\mu(r)$ we took a $\beta$ profile with $\beta=2/3$ (or equivalently
$\mu(r)=\left[1+(x/r_c)^2\right]^{-1.5}$).  For $r_c=1$\arcmin, at small $r$
both profiles differ by 18~\%, and at large radii (also at 15\arcmin\ off-axis)
there is a 10~\% excess.  Even for $r_c=3$\arcmin, there is a significant
effect.  This is all due to the strong psf tails.

The good news, however, is that the ratio $\phi/\mu$ does not vary more than
1~\% (for a core radius of 1\arcmin) when the photon energy is changed from 0 to
9~keV.  This indicates that the psf does not introduce significant spectral
distortions for the clusters that we consider in this paper.  Thus, we
use a correction factor $\phi/\mu$ that is independent of energy, and determine
it from a single energy band.  Therefore the psf correction does not
affect the shape of the spectra, only their normalization.

\subsection{Deprojecting spectra\label{sect:depro}}

We have demonstrated in the previous section that contamination of spectra in
different regions by flux from other regions is an effect that should be taken
into account in the analysis of these spectra.  We describe in the Appendix how
this is formally implemented.  However as shown in that appendix simultaneous
fitting of all spectra in different annuli taking all relevant effects
(vignetting, psf, projection) into account in a forward folding technique is
problematic.  Therefore we correct and deproject our spectra as follows.

We start with a template X-ray image of the cluster.  We have taken for this a
vignetting-corrected Rosat PSPC image as obtained from NASA's skyview facility.
The constant background level was determined at large radii and subtracted from
the image.  In the central parts (typically a few arcmin), where the spatial
variations are large, this PSPC image was merged smoothly with a
background-subtracted Rosat HRI image.  The resulting image was then convolved
with the psf of the EPIC camera's and from this the ratio $\phi(r)/\mu(r)$ cf.
Sect.~\ref{sect:radprof} was determined.

The use of a PSPC image as the template for the true cluster image does not
introduce much bias in our analysis.  This can be seen as follows.  For a
Gaussian psf $F(r) = 1-\exp\left[ -r^2/2\sigma^2\right]$ the ratio
$\phi(r)/\mu(r)$ can be developed in a series as:
\begin{equation}
\label{eqn:mucor_gaus}
\phi(r)/\mu(r) = 1 + 0.5\sigma^2 \left[ \mu^{\prime\prime}(r)
 + \mu^{\prime}(r)/r
          \right]/\mu(r) + O(\sigma^4).
\end{equation}
Therefore the correction factor is of order $(\sigma / a)^2$ with $a$ the
characteristic scale height of $\mu(r)$ (and in the limit of $\sigma\ll a$).
For a non-Gaussian psf similar expressions can be obtained that can be cast in
the shape of (\ref{eqn:mucor_gaus}) with an effective value for $\sigma$ that
depends upon the details of the psf.  The above (\ref{eqn:mucor_gaus}) holds
both for convolution of any radial profile with the Rosat ($\sigma_R$) and
XMM-Newton ($\sigma_X$) psf.  In both cases, $(\sigma / a)^2$ is of the order of
$\sim 10$~\%.  Our estimate for $\phi(r)/\mu(r)$ is $\mu(r)$ convolved with
$\sigma_X$ and $\sigma_R$, dividided by $\mu(r)$ convolved with $\sigma_R$.  The
true value for $\phi(r)/\mu(r)$ is $\mu(r)$ convolved with $\sigma_X$, dividided
by $\mu(r)$.  It is easy to show that the difference between both expressions is
of order $(\sigma_R / a)^2 (\sigma_X / a)^2$, i.e.  on the one percent level.
Therefore there is no need to consider higher resolution template images.

Then for the EPIC data we make a radial, background subtracted profile for each
energy band.  These radial profiles are corrected for exposure (CCD gaps,
removed hot pixels, removed point sources etc.).  We then apply the psf
correction factor $\mu(r)/\phi(r)$, and finally the (energy and position
dependent) vignetting correction in order get the exposure, psf and vignetting
corrected radial profile for each energy band.  Vignetting is both due to the
telescope and in case of the MOS cameras also due to the Reflection Grating
Array.

From these radial profiles we then construct the deprojected count rates in
various shells, assuming spherical symmetry.  It can be shown in general that if
the spatial emissivity is $\epsilon(r)$, the projected surface brightness
$\mu(r)$, and the integrated emissivity $y(r)$ is defined by
\begin{equation}
y(r) \equiv \int\limits_0^{r} 4\pi s^2 \epsilon(s) {\mathrm d}s,
\end{equation}
then $y(r)$ is related to $\mu$ by:
\begin{equation}
\label{eqn:yr}
y(r) = y(\infty) - \int\limits_r^{\infty}
    2\pi t \mu(t) g(t/r) {\mathrm d}t,
\end{equation}
where
\begin{equation}
g(x) \equiv \left\{
\begin{array}{ll}
 0 & \qquad\mbox{if $x < 1$};\\
 \displaystyle{
{2\over \pi \sqrt{x^2-1} } + 1 - {2\over\pi} \arcsin({1\over x})} &
  \qquad\mbox{if $x \geq 1$}.
 \end{array} \right.
\end{equation}

The integration kernel $2\pi t \mu(t){\mathrm d}t$ is proportional to the number
of observed counts in the projected annulus centered around $t$ with width
${\mathrm d}t$.  The difference $Y_{12}\equiv y(r_2)-y(r_1)$ is the number of
{\em emitted} counts in a spherical shell between $r_1$ and $r_2$.  Hence,
(\ref{eqn:yr}) gives the (deprojected) number of emitted counts in a spherical
shell as a linear combination of the observed number of counts in the
(projected) surface brightness profile.  We sample the surface brightness $\mu$
in circular annuli with thickness $\Delta$ and inner and outer radii
$(i-1)\Delta$ and $i\Delta$, and take $N_i$ the number of observed counts in
these projected annuli.  It follows that
\begin{equation}
\label{eqn:y12}
Y_{12} = \sum_{i}^{\,} 
 \left\{ 
  g\Bigl( {\frac{(i+1/2)\Delta}{r_1}} \Bigr) 
- g\Bigl( {\frac{(i+1/2)\Delta}{r_2}} \Bigr)
  \right\}
  N_i.
\end{equation}

Evaluating (\ref{eqn:y12}) for all annuli and energy bins, the deprojected
spectra for each shell are obtained directly.  It is also straightforward to
evaluate the statistical error bars on the spectrum.  In the choice of the shell
boundaries $r_1$ and $r_2$ we have searched a compromise between spatial
resolution and signal to noise ratio.  Therefore we choose the boundary
separations not smaller than 0.5\arcmin\ (for smaller separations, in particular
in the core, psf effects cause strong overlap between the neighbouring shells).
The shell boundaries used in this work are listed in Table~\ref{tab:boundaries}.
These 9 shells cover approximately the entire field of view of the XMM-Newton
telescopes.

\begin{table}[!ht]
\caption{Boundaries between the shells. For each shell nr., the lower
and upper radius (in arcmin) are listed.}
\label{tab:boundaries}
\centerline{
\begin{tabular}{|rc|rc|rc|}
\hline
nr. & $r_1$ -- $r_2$ (\arcmin) &
nr. & $r_1$ -- $r_2$ (\arcmin) &
nr. & $r_1$ -- $r_2$ (\arcmin) \\
\hline
1 & 0 -- 0.5 & 4 & 2 -- 3 & 7 & 6 -- 9 \\
2 & 0.5 -- 1 & 5 & 3 -- 4 & 8 & 9 -- 12 \\
3 & 1 -- 2   & 6 & 4 -- 6 & 9 & 12 -- 16 \\
\hline\noalign{\smallskip}
\end{tabular}
}
\end{table}

A possible complication arises for those clusters that have X-ray emission
outside the field of view of the XMM-Newton telescopes.  We have not attempted
to estimate a correction for this in (\ref{eqn:y12}).  For all our clusters, the
surface brightness drops rapidly towards large radii.  Therefore the projected
contribution from shells outside the field of view to the X-ray flux of the
innermost shells with cooling gas is negligible.  In the outer parts of the
clusters, the shape of the X-ray spectrum does not vary rapidly with radius, so
neglecting the out-of-field emission only gives a small bias to the
normalisation of the spectra of the outermost shells, not to their shape.  But
as noted, several of our clusters have no significant or only a small flux
outside 16\arcmin\ radius.  However, for safety we do not consider here the
spectra of the outermost annulus 9 (12\arcmin--16\arcmin), although the counts
in this region have been taken into account in the deprojection for the inner
shells.

Finally, we note that since all correction factors (vignetting, exposure, psf,
projection) are taken into account in the deprojected spectra, we can use the
nominal, on-axis response matrices for the spectral analysis.  For these we took
the v9q20 precanned matrices from the XMM-Newton calibration database (version
April/May 2002).

\subsection{Projected spectra}

Deprojecting spectra is absolutely necessary for studying the innermost,
cooling parts of the cluster. For proper modeling, any local hot gas
in the core must be separated from projected hot gas from the outer parts.
This is evident if one considers for example the strong abundance
gradients observed in several clusters (Kaastra et al. \cite{kaastra01};
Tamura et al. \cite{tamura01a}, \cite{tamura01b}). However, the deprojection
increases slightly the noise in the spectra, because also counts
from the outermost annuli are taken into account in arriving at the
deprojected spectrum. For studies of the outer parts of the cluster
(for example the mass profile at large radii) deprojection is less urgent
due to the slow variations of the spectral parameters with radius in those
regions.

In the present work we do not use the non-deprojected spectra, however in a
subsequent paper (Tamura et al.  \cite{tamura03}, on abundances in clusters) we
use extensively the non-deprojected spectra; since in that paper we will not
repeat extensively all these analysis issues, we give them here for
completeness.

When we use these non-deprojected spectra, we follow a more straightforward
approach.  We start with the template image (a Rosat PSPC image), apply the
vignetting correction, convolve it explicitly with the instrumental psf and
apply the exposure corrections.  We then determine for each annulus the ratio of
the number of PSPC counts in this corrected image to the number of PSPC counts
in the original template image, and this determines the relative effective area
correction (arf) that we apply to the standard response matrix.  This is
essentially the same method as used in the analysis of BeppoSAX MECS data of
clusters by D'Acri et al.  (\cite{acri98}).

\subsection{Spectral binning}

The spectra for each annulus were accumulated in 15~eV bins.  These bins were
then rebinned further onto a grid with a spacing of about 1/3~FWHM for all three
detectors.  We used the same energy grid for all three detectors.  The data were
binned then further by a factor between 2--12 as listed in Table~\ref{tab:ebin}.
This was done in order to enhance the sensitivity in parts of the spectrum where
there is only continuum or with very weak lines.  It enhances the S/N ratio for
the weak spectra of the outer parts.  Spectral fitting was restricted to the
0.2--10~keV range.  The energy range below 0.2~keV is currently too poorly
calibrated to be useful for spectral analysis, while above 10~keV most of our
spectra lack sufficient flux.

\begin{table}[!ht]
\caption{Spectral binning. Number of bins of 1/3 FWHM taken together
in intervals $E_1$--$E_2$, in the rest frame of the cluster.}
\label{tab:ebin}
\centerline{
\begin{tabular}{|rrrl|}
\hline
$E_1$ & $E_2$ & binning & comments\\
(keV) & (keV) & factor  & \\
\hline
0.20 & 0.50 &  2 & most continuum\\
0.50 & 3.50 &  2 & many densely spaced lines\\
3.50 & 3.75 &  8 & continuum between Ar and Ca\\
3.75 & 4.25 &  3 & Ca lines\\
4.25 & 6.30 & 12 & continuum between Ca and Fe\\
6.30 & 7.20 &  4 & Fe-K lines\\
7.20 & 7.50 &  8 & continuum\\
7.50 & 8.70 &  4 & Fe-K, Ni-K lines\\
8.70 & 10.00 & 8 & continuum\\
\hline\noalign{\smallskip}
\end{tabular}
}
\end{table}

For the spectral analysis, we used the SPEX package (Kaastra et al.
\cite{kaastra02spex}).  For the interstellar absorption, we used the Morrison \&
McCammon (\cite{morrison83}) model, and for the thermal plasma emission we used
the collisional ionisation model (CIE) as available in SPEX.  Systematic errors,
both as a fraction of the source spectrum and the background spectrum, were
added according to the prescriptions of the next section.  The spectra of the
three camera's were fit simultaneously, but in our figures we added them for
clarity.

\subsection{Systematic uncertainties}

There are remaining systematic uncertainties in both the effective area to be
used in the spectral response matrices, as well as in the subtracted background
level.  Based upon a study of several calibration sources by us, we have put
conservatively systematic errors to the spectrum of 10~\% of the source flux
below 0.3~keV and above 2~keV, and 5~\% in the 0.3--2~keV range.

In addition we included a systematic background error of 35~\% of the total
background below 0.5~keV, 25~\% between 0.5--0.7~keV, 15~\% between 0.7--2~keV,
and 10~\% above 10~keV.  This was based upon our discussion in
Sect.~\ref{sect:back}.  Both systematic errors were added in quadrature to the
binned spectra.

Finally, there is still a systematic difference in normalisation between the
three EPIC camera's.  Therefore in our spectral fits we allowed the relative
normalisation of MOS2 and pn with respect to MOS1 to vary.  In our fits, we
found no systematic dependency of these renormalisations as a function of
radius.  For our full cluster sample, the average MOS2/MOS1 renormalisation
factor was $1.011\pm 0.014$, while the pn/MOS1 renormalisation factor was
$0.898\pm 0.013$.  All spectral normalisations in the present work are related
to MOS1.

\section{Spectral modeling\label{sect:results}}

\subsection{Spectral model}

There are several ways to model the spectrum of cooling gas in clusters of
galaxies.  These include multi-temperature fitting, parameter fitting of
isobaric cooling flow models, both with and without the inclusion of an
additional hot gas component, etc.  The key ingredient in all these models is a
prescription of the differential emission measure distribution ${\rm d}Y(T)/{\rm
d}T$, where the emission measure $Y$ is the volume integral of the product
$n_{\mathrm e}n_{\mathrm H}$, with $n_{\mathrm e}$ the electron density and
$n_{\mathrm H}$ the hydrogen density.

In the present paper we attempt to do the spectral fitting as much as possible
independently of any assumed physical model for ${\rm d}Y(T)/{\rm d}T$.  Based
upon our spectral fits we make a numerical representation of ${\rm d}Y(T)/{\rm
d}T$.  This numerical representation can be compared afterwards with any model
for the cooling gas.  Therefore, a kind of multi-temperature fitting is the
preferred way to proceed.

In practice, the best temperature diagnostics for clusters of galaxies are
obtained from the exponential turnover of the Bremsstrahlung continuum, in
particular for the hottest gas, and the line emission from mainly the Fe-L and
Fe-K complexes.  In particular the Fe-L complex is a good temperature indicator
for the cooler gas in the cluster.  The line power emitted by any ion peaks at a
characteristic temperature, and has a typical FWHM of the order of a factor of
two in temperature.  At each temperature, there are usually only a few ions with
large emissivity.  This combined with the factor of two FWHM for each ion makes
it useless to sample ${\rm d}Y(T)/{\rm d}T$ on a temperature grid that is finer
than $\sim$0.15 in $\log T$.  We performed simulations and found that a
temperature binning of 0.15 in $\log T$ is sufficient to reconstruct any cluster
spectrum, as compared to an arbitrary continuous or discrete emission measure
distribution.  However, with a spacing of 0.15 in $\log T$ we sometimes found
strong correlations between the emission measure values of neighbouring bins.
The reason is that the isothermal spectrum of a given temperature component can
be approximated by a linear combination of the spectra of its neighbouring
temperature bins.  When the observed spectrum is noisy, it is hard to see the
difference, and as a result strong correlations between the components arise.
We found that with a spacing of 0.30 in $\log T$ these correlations disappear,
while the temperature resolution is still sufficient to reconstruct ${\rm
d}Y(T)/{\rm d}T$ completely.

Therefore we have used a multi-temperature CIE model with temperatures $T_1$,
$T_2$, etc., emission measures $Y_1$, $Y_2$, etc., constrained by the condition
that $T_{i+1} = 0.5 T_i$.  We leave the temperature of the hottest component,
$T_1$, a free parameter, in order not to distribute the emission measure over
two bins in case the spectrum is isothermal.  This multi-temperature CIE model
was redshifted with the cosmological redshift of each cluster and absorbed by
Galactic absorption.  Redshifts and column densities are given in
Table~\ref{tab:cluspar}.  We have not found evidence for significant absorption
above the Galactic values; this is confirmed by the higher resolution RGS data
for the same clusters (Peterson et al.  \cite{peterson03}); but even if there
were additional absorption, this would not affect our multi-temperature fitting
too much, since the strongest effect of absorption is the depth of the oxygen
edge, and this edge energy is well below the energy of most of the strong Fe-L
lines.

In the present paper we focus upon the temperature structure; as we explained
the Fe-L complex is a good temperature indicator.  Therefore we left the
abundance of Fe free.  Since K-shell lines from Ne and Mg blend with Fe-L lines,
we also left the Ne and Mg abundance free.  Both O and Si produce relatively
strong spectral lines, and therefore we also left their abundances free.  The
other elements produce weaker lines that do not influence the emission measure
estimate; since we are not interested in abundances in this paper, we couple the
abundances of C and N to O, and those of Na, Al, S, Ar, Ca and Ni to Si.  We
assumed that all CIE components have the same abundance.  The abundances are
studied in greater detail by Tamura et al.  (\cite{tamura03}).

In order to keep the fitting procedure stable and reproducible, we started each
time with a single temperature component with fixed temperature and abundances.
In subsequent steps the number of free parameters is increased, and each time a
spectral fit is made, starting with the most sensitive parameters.  These
parameters are in order:  the temperature of the hot component, its iron
abundance, its other abundances, the renormalisation between the different
instruments.  Then one by one the cooler temperature components were added.  As
a last step, the error bars for the relevant parameters were determined.

\subsection{Single-temperature fitting}

To facilitate comparison with previous work, we first present the results for
single temperature fitting (Table~\ref{tab:fit1t}).  In general, we give
temperatures and densities for the innermost 8 shells, except when the data
become too noisy in the outermost shell, or in the innermost shell whenever
there is a known strong X-ray emitting AGN in the center (Virgo, Perseus) or the
data are noisy due to low surface brightness in combination with small
extraction radius (A~3266).  In Coma, we omit the innermost shell since it is
strongly contaminated by NGC~4874, which we adopted as the center of our annuli.
Since we study cooling flows in this paper, we also restrict the radial range to
cooling times less than 100 Gyr.

The temperature profiles that we derive for the single-temperature fits are in
good agreement with what we or others derived before from measurements with
other satellites or from the same XMM-Newton data
(M~87: B\"ohringer et al. \cite{boehringer01},
Matsushita et al. \cite{matsushita02}; 
S\'ersic~159$-$3: Kaastra et al. \cite{kaastra01};
A~496: Tamura et al. \cite{tamura01b};
A~1795: Tamura et al. \cite{tamura01a}, Arnaud et al. \cite{arnaud01};
A~1835: Majerowicz et al. \cite{majerowicz02}).

Although the formal statistical error bars on the temperature and density are
often very small, this does not prove that the spectra at each radius are
strictly isothermal.  The small error bars merely indicate the high statistical
quality of our spectra.  In fact, in particular in the inner parts of many
clusters the $\chi^2$ of the fits are statistically not acceptable, contrary to
the multi-temperature fits of Sect.~\ref{sect:multit} below.

\begin{table*}[!ht]
\caption{Spectral fit results for a single-temperature model. The \#\ indicates
the shell number (cf. Table~\ref{tab:boundaries}).}
\label{tab:fit1t}
\centerline{
\begin{tabular}{|l|rrrr|rrrr|rrrr|}
\hline
\# &
 $\chi^2$  & $kT$ & $\log n_{\mathrm H}$  & $\tau_{\mathrm{\hbox{cool}}}$ & 
 $\chi^2$  & $kT$ & $\log n_{\mathrm H}$  & $\tau_{\mathrm{\hbox{cool}}}$ & 
 $\chi^2$  & $kT$ & $\log n_{\mathrm H}$  & $\tau_{\mathrm{\hbox{cool}}}$ \\
   & 
  &(keV)&   (m$^{-3}$)          & (Gyr) &
  &(keV)&   (m$^{-3}$)          & (Gyr) &
  &(keV)&   (m$^{-3}$)          & (Gyr) \\
\hline
\multicolumn{1}{|l|}{ }&
\multicolumn{4}{|c|}{NGC 533 (dof=237)}&
\multicolumn{4}{|c|}{Virgo (dof=237)}&
\multicolumn{4}{|c|}{A 262 (dof=237)}\\
\hline
1& 393& 0.67$\pm$ 0.01 &  3.73$\pm$ 0.03 &   0.8 &    &                &                 &       & 487& 1.02$\pm$ 0.02 &  4.05$\pm$ 0.02 &   0.7\\
2& 348& 0.96$\pm$ 0.03 &  3.17$\pm$ 0.04 &   4.6 & 635& 1.45$\pm$ 0.02 &  4.48$\pm$ 0.01 &   0.4 & 241& 1.41$\pm$ 0.03 &  3.71$\pm$ 0.02 &   2.3\\
3& 248& 1.23$\pm$ 0.04 &  2.85$\pm$ 0.03 &    14 & 494& 1.48$\pm$ 0.02 &  4.14$\pm$ 0.01 &   0.9 & 257& 1.88$\pm$ 0.08 &  3.38$\pm$ 0.02 &   7.2\\
4& 271& 1.35$\pm$ 0.11 &  2.66$\pm$ 0.03 &    25 & 338& 1.83$\pm$ 0.03 &  3.97$\pm$ 0.01 &   1.8 & 233& 2.13$\pm$ 0.11 &  3.21$\pm$ 0.02 &    12\\
5& 266& 1.25$\pm$ 0.22 &  2.53$\pm$ 0.05 &    30 & 271& 1.89$\pm$ 0.04 &  3.83$\pm$ 0.01 &   2.5 & 264& 2.16$\pm$ 0.15 &  3.07$\pm$ 0.02 &    17\\
6& 244& 1.18$\pm$ 0.07 &  2.34$\pm$ 0.04 &    42 & 328& 2.06$\pm$ 0.04 &  3.66$\pm$ 0.01 &   4.1 & 268& 2.16$\pm$ 0.10 &  2.93$\pm$ 0.01 &    23\\
7& 221& 1.00$\pm$ 0.11 &  2.24$\pm$ 0.04 &    42 & 283& 2.34$\pm$ 0.04 &  3.50$\pm$ 0.01 &   6.5 & 244& 2.25$\pm$ 0.12 &  2.71$\pm$ 0.01 &    39\\
8&    & 	       &		 &	 & 258& 2.50$\pm$ 0.05 &  3.37$\pm$ 0.00 &   9.4 & 213& 2.05$\pm$ 0.16 &  2.56$\pm$ 0.02 &    52\\
\hline
\multicolumn{1}{|l|}{ }&
\multicolumn{4}{|c|}{A 1837 (dof=237)}&
\multicolumn{4}{|c|}{S\'ersic 159$-$3 (dof=237)}&
\multicolumn{4}{|c|}{MKW 9 (dof=237)}\\
\hline
1& 232& 2.93$\pm$ 0.20 &  3.47$\pm$ 0.03 &   8.4 & 288& 2.17$\pm$ 0.05 &  4.12$\pm$ 0.01 &   1.5 & 300& 1.28$\pm$ 0.09 &  3.31$\pm$ 0.07 &   5.1\\
2& 261& 4.61$\pm$ 0.52 &  3.14$\pm$ 0.03 &    24 & 288& 2.37$\pm$ 0.05 &  3.77$\pm$ 0.01 &   3.5 & 228& 1.66$\pm$ 0.20 &  3.08$\pm$ 0.05 &    12\\
3& 258& 4.15$\pm$ 0.22 &  2.96$\pm$ 0.01 &    34 & 278& 2.43$\pm$ 0.05 &  3.35$\pm$ 0.01 &   9.6 & 219& 2.25$\pm$ 0.27 &  2.84$\pm$ 0.03 &    29\\
4& 274& 3.65$\pm$ 0.22 &  2.71$\pm$ 0.02 &    55 & 275& 2.38$\pm$ 0.10 &  2.98$\pm$ 0.01 &    22 & 239& 2.45$\pm$ 0.37 &  2.58$\pm$ 0.05 &    57\\
5& 248& 3.57$\pm$ 0.33 &  2.54$\pm$ 0.02 &    82 & 270& 2.43$\pm$ 0.16 &  2.73$\pm$ 0.01 &    40 & 231& 2.06$\pm$ 0.72 &  2.44$\pm$ 0.05 &    67\\
6&    & 	       &		 &	 & 244& 2.07$\pm$ 0.16 &  2.48$\pm$ 0.01 &    62 & 281& 1.60$\pm$ 0.33 &  2.37$\pm$ 0.04 &    60\\
\hline
\multicolumn{1}{|l|}{ }&
\multicolumn{4}{|c|}{2A 0335+096 (dof=399)}&
\multicolumn{4}{|c|}{MKW 3s (dof=237)}&
\multicolumn{4}{|c|}{A 2052 (dof=237)}\\
\hline
1& 585& 1.40$\pm$ 0.03 &  4.38$\pm$ 0.02 &   0.5 & 245& 3.00$\pm$ 0.10 &  3.93$\pm$ 0.01 &   2.9 & 554& 1.41$\pm$ 0.07 &  4.07$\pm$ 0.01 &   1.0\\
2& 480& 1.87$\pm$ 0.04 &  4.06$\pm$ 0.01 &   1.5 & 250& 3.33$\pm$ 0.10 &  3.70$\pm$ 0.01 &   5.4 & 274& 2.54$\pm$ 0.06 &  3.80$\pm$ 0.01 &   3.5\\
3& 450& 2.44$\pm$ 0.06 &  3.64$\pm$ 0.01 &   4.9 & 239& 3.39$\pm$ 0.08 &  3.37$\pm$ 0.01 &    12 & 233& 2.92$\pm$ 0.07 &  3.40$\pm$ 0.01 &   9.7\\
4& 469& 2.86$\pm$ 0.12 &  3.29$\pm$ 0.02 &    12 & 266& 3.79$\pm$ 0.13 &  3.10$\pm$ 0.01 &    23 & 258& 3.08$\pm$ 0.10 &  3.16$\pm$ 0.01 &    18\\
5& 425& 3.02$\pm$ 0.18 &  3.11$\pm$ 0.03 &    20 & 257& 3.43$\pm$ 0.17 &  2.89$\pm$ 0.01 &    36 & 235& 3.16$\pm$ 0.15 &  2.97$\pm$ 0.01 &    28\\
6& 452& 2.86$\pm$ 0.18 &  2.87$\pm$ 0.02 &    33 & 201& 3.44$\pm$ 0.17 &  2.66$\pm$ 0.01 &    61 & 242& 3.06$\pm$ 0.12 &  2.77$\pm$ 0.01 &    43\\
7& 377& 2.68$\pm$ 0.26 &  2.55$\pm$ 0.03 &    65 & 301& 2.54$\pm$ 0.17 &  2.38$\pm$ 0.01 &    93 & 276& 2.55$\pm$ 0.12 &  2.50$\pm$ 0.01 &    70\\
8&    &                &                 &       & 370& 1.70$\pm$ 0.13 &  2.26$\pm$ 0.02 &    84 & 322& 1.54$\pm$ 0.13 &  2.32$\pm$ 0.02 &    65\\
\hline
\multicolumn{1}{|l|}{ }&
\multicolumn{4}{|c|}{A 4059 (dof=399)}&
\multicolumn{4}{|c|}{Hydra A (dof=237)}&
\multicolumn{4}{|c|}{A 496 (dof=237)}\\
\hline
1& 467& 2.11$\pm$ 0.14 &  3.81$\pm$ 0.02 &   3.0 & 288& 2.92$\pm$ 0.10 &  4.18$\pm$ 0.01 &   1.6 & 271& 2.14$\pm$ 0.10 &  4.17$\pm$ 0.02 &   1.3\\
2& 402& 3.12$\pm$ 0.15 &  3.58$\pm$ 0.02 &   6.8 & 253& 3.32$\pm$ 0.12 &  3.83$\pm$ 0.01 &   4.0 & 244& 2.51$\pm$ 0.12 &  3.86$\pm$ 0.01 &   3.0\\
3& 399& 3.90$\pm$ 0.12 &  3.33$\pm$ 0.01 &    14 & 243& 3.19$\pm$ 0.12 &  3.38$\pm$ 0.01 &    11 & 278& 3.34$\pm$ 0.13 &  3.52$\pm$ 0.01 &   8.2\\
4& 378& 3.89$\pm$ 0.19 &  3.09$\pm$ 0.01 &    24 & 265& 3.28$\pm$ 0.17 &  3.16$\pm$ 0.01 &    18 & 272& 3.93$\pm$ 0.24 &  3.25$\pm$ 0.02 &    17\\
5& 410& 4.14$\pm$ 0.26 &  2.86$\pm$ 0.01 &    43 & 236& 3.59$\pm$ 0.25 &  2.95$\pm$ 0.01 &    32 & 172& 4.03$\pm$ 0.30 &  3.06$\pm$ 0.02 &    26\\
6& 388& 4.26$\pm$ 0.30 &  2.65$\pm$ 0.01 &    71 & 268& 3.38$\pm$ 0.27 &  2.64$\pm$ 0.02 &    62 & 231& 4.63$\pm$ 0.36 &  2.88$\pm$ 0.02 &    44\\
7&    &                &                 &       &    & 	       &		 &	 & 254& 4.02$\pm$ 0.34 &  2.64$\pm$ 0.02 &    70\\
\hline
\multicolumn{1}{|l|}{ }&
\multicolumn{4}{|c|}{A 3112 (dof=237)}&
\multicolumn{4}{|c|}{A 1795 (dof=237)}&
\multicolumn{4}{|c|}{A 399 (dof=237)}\\
\hline
1& 317& 2.99$\pm$ 0.08 &  4.10$\pm$ 0.01 &   2.0 & 245& 3.47$\pm$ 0.09 &  4.13$\pm$ 0.01 &   2.1 & 285& 2.60$\pm$ 0.58 &  3.38$\pm$ 0.07 &   9.3\\
2& 286& 3.96$\pm$ 0.14 &  3.69$\pm$ 0.01 &   6.2 & 225& 4.15$\pm$ 0.10 &  3.85$\pm$ 0.01 &   4.3 & 239& 4.63$\pm$ 0.84 &  3.31$\pm$ 0.03 &    16\\
3& 381& 4.34$\pm$ 0.12 &  3.34$\pm$ 0.01 &    15 & 229& 5.23$\pm$ 0.14 &  3.47$\pm$ 0.01 &    12 & 210& 5.89$\pm$ 0.71 &  3.05$\pm$ 0.03 &    34\\
4& 267& 4.42$\pm$ 0.19 &  2.99$\pm$ 0.01 &    33 & 188& 5.56$\pm$ 0.21 &  3.19$\pm$ 0.01 &    24 & 206& 7.78$\pm$ 1.18 &  2.92$\pm$ 0.03 &    54\\
5& 304& 3.78$\pm$ 0.34 &  2.73$\pm$ 0.02 &    55 & 195& 5.76$\pm$ 0.29 &  2.97$\pm$ 0.01 &    40 & 240& 6.23$\pm$ 0.88 &  2.77$\pm$ 0.03 &    68\\
6&    &                &                 &       & 279& 5.89$\pm$ 0.35 &  2.72$\pm$ 0.01 &    73 &    & 	       &		 &	\\
\hline
\multicolumn{1}{|l|}{ }&
\multicolumn{4}{|c|}{A 3266 (dof=237)}&
\multicolumn{4}{|c|}{Perseus (dof=237)}&
\multicolumn{4}{|c|}{Coma (dof=237)}\\
\hline
1&    & 	       &		 &	 & 624& 5.35$\pm$ 0.16 &  4.55$\pm$ 0.01 &   1.0 &    &                &                 &      \\
2& 299& 7.69$\pm$ 1.68 &  3.38$\pm$ 0.04 &    19 & 393& 3.07$\pm$ 0.05 &  4.32$\pm$ 0.01 &   1.2 & 244&11.19$\pm$ 6.40 &  3.36$\pm$ 0.03 &    24\\
3& 249& 8.35$\pm$ 1.00 &  3.09$\pm$ 0.02 &    38 & 626& 3.21$\pm$ 0.03 &  4.11$\pm$ 0.00 &   2.0 & 230& 6.45$\pm$ 1.12 &  3.21$\pm$ 0.03 &    25\\
4& 231& 9.81$\pm$ 1.78 &  2.92$\pm$ 0.01 &    62 & 392& 3.58$\pm$ 0.05 &  3.90$\pm$ 0.00 &   3.5 & 218& 7.86$\pm$ 1.14 &  3.15$\pm$ 0.02 &    33\\
5& 217& 9.99$\pm$ 1.77 &  2.82$\pm$ 0.02 &    80 & 332& 4.34$\pm$ 0.11 &  3.68$\pm$ 0.01 &   6.6 & 227& 6.92$\pm$ 0.87 &  3.09$\pm$ 0.02 &    34\\
6& 219& 8.37$\pm$ 0.73 &  2.70$\pm$ 0.01 &    94 & 252& 4.97$\pm$ 0.15 &  3.45$\pm$ 0.01 &    12 & 272& 7.80$\pm$ 0.57 &  3.07$\pm$ 0.01 &    39\\
7&    & 	       &		 &	 & 248& 5.27$\pm$ 0.23 &  3.18$\pm$ 0.01 &    24 & 205& 7.73$\pm$ 0.38 &  2.96$\pm$ 0.01 &    50\\
8&    & 	       &		 &	 & 200& 5.45$\pm$ 0.37 &  3.03$\pm$ 0.01 &    34 & 233& 7.21$\pm$ 0.40 &  2.83$\pm$ 0.01 &    64\\
\hline
\multicolumn{1}{|l|}{ }&
\multicolumn{4}{|c|}{A 754 (dof=237)}&
\multicolumn{4}{|c|}{A 1835 (dof=156)}&
\multicolumn{4}{|c|}{ }\\
\hline
1& 280& 5.21$\pm$ 3.04 &  3.17$\pm$ 0.21 &    24 & 185& 5.13$\pm$ 0.14 &  4.12$\pm$ 0.01 &   2.7 &&&&\\
2& 232& 6.18$\pm$ 1.29 &  3.34$\pm$ 0.02 &    18 & 179& 6.91$\pm$ 0.43 &  3.57$\pm$ 0.01 &    11 &&&&\\
3& 238& 6.26$\pm$ 0.55 &  3.21$\pm$ 0.02 &    25 & 178& 7.25$\pm$ 0.47 &  3.08$\pm$ 0.01 &    36 &&&&\\
4& 241& 7.78$\pm$ 0.84 &  3.03$\pm$ 0.02 &    42 & 166& 7.01$\pm$ 0.86 &  2.67$\pm$ 0.02 &    91 &&&&\\
5& 270& 7.76$\pm$ 1.04 &  2.90$\pm$ 0.02 &    56 &    & 	       &		 &	 &&&&\\
6& 229& 8.30$\pm$ 0.83 &  2.72$\pm$ 0.02 &    90 &    & 	       &		 &	 &&&&\\
\hline\noalign{\smallskip}
\end{tabular}
}
\end{table*}

\subsection{Multi-temperature fitting\label{sect:multit}}

We list our results obtained using the multi-temperature fitting described in
the previous section in Tables~\ref{tab:fitpar1}--\ref{tab:fitpar2}.  We only
list parameters for shells with a cooling time less than 30 Gyr, as determined
from our single-temperature fits of the previous section.  For most clusters the
density outside that radius is low and as a result the X-ray flux is low,
resulting in a spectrum that is often too noisy for multi-temperature fitting.

We give the root-mean-square error bars on all emission measures
($\Delta\chi^2=2$).  We averaged the positive and negative error bars:  in most
cases the differences between these errors is not very large.  In those cases
where the best-fit emission measure is zero, we only list the upper error bar:
our model does not allow for negative emission measures for obvious reasons.

\begin{figure}
\resizebox{\hsize}{!}{\includegraphics[angle=-90]{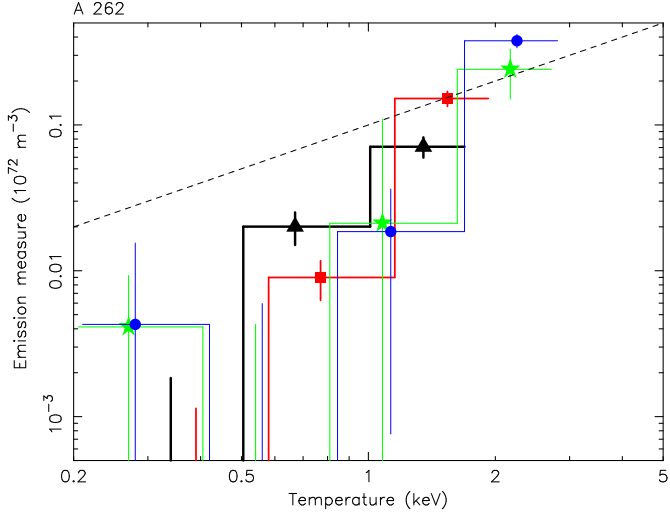}}
\caption{Differential emission measure distribution for A~262 in the innermost
four shells. Line width decreases with radius; symbols: triangles:
0--0.5\arcmin; squares: 0.5--1\arcmin; stars: 1--2\arcmin; circles:
2--3\arcmin. The dashed line approximates the slope of the isobaric cooling
flow model.}
\label{fig:cf_a262}
\end{figure}

In Fig.~\ref{fig:cf_a262} we plot as a typical example the differential emission
measure distribution for A~262 in the innermost four shells (where the radiative
cooling time is shorter than the Hubble time).  It is seen that the spectra are
not strictly isothermal:  there is a significant emission measure contribution
at $T_2=0.5T_1$, although the emission measure drops rapidly towards lower
temperatures.  Note also that $T_1$ increases with radius.  Finally, the lowest
temperature in an outer shell extends lower than the highest temperature in an
inner shell.  So we are not just seeing a temperature gradient smoothed in a
given shell.  The properties described in this paragraph for A~262 are typical
for all the cooling clusters in our sample.

\begin{table*}[!ht]
\caption{Spectral fit results for the multi-temperature model.
The temperature is the temperature of the hottest
component, $T_1$; the emission measures $Y_1$ to $Y_5$ have been renormalised
to 1 for $Y_1$. The \#\ indicates the shell number. A "C" in the last column
indicates consistency with an isobaric cooling flow spectrum, an "I"
consistency with an isothermal spectrum.
}
\label{tab:fitpar1}
\centerline{
\begin{tabular}{|lrrrrr|rrrrr|r|}
\hline
\# & $\chi^2$ & d.o.f. & $kT_1$ & $\log n_{\mathrm H}$  & $\tau_{\mathrm{\hbox{cool}}}$ 
  & $Y_{1}$ &$Y_{2}$ &$Y_{3}$ &$Y_{4}$ &$Y_{5}$ &  \\
   &          &        & (keV)&   (m$^{-3}$)          & (Gyr) &&&&&& \\
\hline
\multicolumn{12}{|l|}{NGC 533}\\
\hline
1& 393& 233& 0.7$^{+ 0.0}_{-0.0}$& 3.73&  0.8&1$\pm$ 0.14 & 0.00$\pm$ 0.44 & 0.00$\pm$ 0.01 & 0.00$\pm$ 0.03 & 0.00$\pm$ 0.59&I\\
2& 297& 233& 1.1$^{+ 0.1}_{-0.0}$& 3.10&  6.6&1$\pm$ 0.22 & 0.19$\pm$ 0.29 & 0.00$\pm$ 0.02 & 0.00$\pm$ 0.03 & 0.00$\pm$ 0.13&CI\\
3& 210& 233& 1.9$^{+ 0.4}_{-0.4}$& 2.73&   32&1$\pm$ 0.32 & 0.28$\pm$ 0.21 & 0.02$\pm$ 0.04 & 0.00$\pm$ 0.06 & 0.00$\pm$ 0.05&I\\
4& 269& 233& 1.4$^{+ 0.2}_{-0.1}$& 2.62&   30&1$\pm$ 0.29 & 0.01$\pm$ 0.13 & 0.07$\pm$ 0.07 & 0.05$\pm$ 0.22 & 0.00$\pm$ 0.64&I\\
5& 263& 233& 1.5$^{+ 0.4}_{-0.3}$& 2.47&   46&1$\pm$ 0.34 & 0.15$\pm$ 0.12 & 0.00$\pm$ 0.08 & 0.00$\pm$ 0.37 & 0.01$\pm$ 1.62&I\\
\hline
\multicolumn{12}{|l|}{Virgo}\\
\hline
2& 357& 233& 1.7$^{+ 0.0}_{-0.0}$& 4.41&  0.6&1$\pm$ 0.05 & 0.11$\pm$ 0.01 & 0.0$\pm$ 0.003 & 0.00$\pm$ 0.01 & 0.05$\pm$ 0.02&\\
3& 325& 233& 1.7$^{+ 0.0}_{-0.0}$& 4.09&  1.2&1$\pm$ 0.05 & 0.07$\pm$ 0.01 & 0.0$\pm$ 0.002 & 0.00$\pm$ 0.01 & 0.05$\pm$ 0.02&\\
4& 268& 233& 2.0$^{+ 0.1}_{ 0.0}$& 3.93&  2.1&1$\pm$ 0.05 & 0.07$\pm$ 0.01 & 0.0$\pm$ 0.002 & 0.00$\pm$ 0.01 & 0.02$\pm$ 0.01&\\
5& 239& 233& 2.0$^{+ 0.1}_{-0.1}$& 3.80&  2.9&1$\pm$ 0.06 & 0.05$\pm$ 0.02 & 0.0$\pm$ 0.004 & 0.00$\pm$ 0.02 & 0.02$\pm$ 0.02&\\
6& 277& 233& 2.3$^{+ 0.8}_{-0.1}$& 3.62&  5.0&1$\pm$ 0.25 & 0.13$\pm$ 0.24 & 0.0$\pm$ 0.043 & 0.00$\pm$ 0.01 & 0.04$\pm$ 0.01&I\\
7& 263& 233& 2.6$^{+ 1.6}_{-0.2}$& 3.47&  7.8&1$\pm$ 0.34 & 0.12$\pm$ 0.29 & 0.0$\pm$ 0.004 & 0.02$\pm$ 0.01 & 0.04$\pm$ 0.02&I\\
8& 238& 233& 3.0$^{+ 0.5}_{-0.3}$& 3.30&   13&1$\pm$ 0.25 & 0.35$\pm$ 0.27 & 0.0$\pm$ 0.009 & 0.00$\pm$ 0.01 & 0.05$\pm$ 0.03&I\\
\hline
\multicolumn{12}{|l|}{A 262}\\		
\hline
1& 327& 233& 1.3$^{+ 0.1}_{-0.1}$& 3.92&  1.3&1$\pm$ 0.16 & 0.28$\pm$ 0.07 & 0.00$\pm$ 0.03 & 0.00$\pm$ 0.05 & 0.00$\pm$ 0.20&\\
2& 220& 233& 1.5$^{+ 0.2}_{-0.1}$& 3.67&  2.9&1$\pm$ 0.12 & 0.06$\pm$ 0.02 & 0.00$\pm$ 0.01 & 0.00$\pm$ 0.02 & 0.00$\pm$ 0.07&\\
3& 236& 233& 2.2$^{+ 1.3}_{-0.1}$& 3.31&  9.5&1$\pm$ 0.37 & 0.09$\pm$ 0.37 & 0.00$\pm$ 0.02 & 0.02$\pm$ 0.02 & 0.06$\pm$ 0.04&I\\
4& 230& 233& 2.3$^{+ 0.2}_{-0.1}$& 3.19&   13&1$\pm$ 0.10 & 0.05$\pm$ 0.05 & 0.00$\pm$ 0.02 & 0.01$\pm$ 0.03 & 0.02$\pm$ 0.05&I\\
5& 255& 233& 3.5$^{+ 2.8}_{-1.1}$& 2.86&   38&1$\pm$ 1.25 & 1.32$\pm$ 1.04 & 0.08$\pm$ 0.51 & 0.00$\pm$ 0.10 & 0.03$\pm$ 0.19&CI\\
6& 262& 233& 2.4$^{+ 0.3}_{-0.2}$& 2.91&   26&1$\pm$ 0.13 & 0.05$\pm$ 0.12 & 0.00$\pm$ 0.03 & 0.04$\pm$ 0.03 & 0.09$\pm$ 0.11&I\\
\hline
\multicolumn{12}{|l|}{A 1837}\\
\hline
1& 226& 233& 3.1$^{+ 0.2}_{-0.2}$& 3.44&  9.3&1$\pm$ 0.15 & 0.00$\pm$ 0.28 & 0.01$\pm$ 0.01 & 0.00$\pm$ 0.01 & 0.03$\pm$ 0.03&I\\
2& 254& 233& 7.3$^{+ 0.7}_{-2.0}$& 3.07&   37&1$\pm$ 0.24 & 0.00$\pm$ 0.47 & 0.41$\pm$ 0.26 & 0.01$\pm$ 0.19 & 0.03$\pm$ 0.05&C\\
3& 253& 233& 6.6$^{+ 1.4}_{-2.5}$& 2.79&   67&1$\pm$ 1.05 & 1.14$\pm$ 0.71 & 0.11$\pm$ 0.89 & 0.02$\pm$ 0.05 & 0.02$\pm$ 0.12&CI\\
\hline
\multicolumn{12}{|l|}{S\'ersic~159$-$3}\\
\hline
1& 265& 233& 2.3$^{+ 0.1}_{-0.1}$& 4.09&  1.7&1$\pm$ 0.06 & 0.07$\pm$ 0.03 & 0.00$\pm$ 0.01 & 0.02$\pm$ 0.01 & 0.10$\pm$ 0.04&\\
2& 203& 233& 2.9$^{+ 0.2}_{-0.2}$& 3.70&  4.8&1$\pm$ 0.12 & 0.28$\pm$ 0.12 & 0.00$\pm$ 0.01 & 0.00$\pm$ 0.01 & 0.31$\pm$ 0.07&I\\
3& 203& 233& 2.6$^{+ 0.1}_{-0.1}$& 3.33&   10&1$\pm$ 0.04 & 0.00$\pm$ 0.39 & 0.00$\pm$ 0.01 & 0.01$\pm$ 0.01 & 0.25$\pm$ 0.07&I\\
4& 242& 233& 2.7$^{+ 0.1}_{-0.1}$& 2.96&   25&1$\pm$ 0.06 & 0.00$\pm$ 0.28 & 0.00$\pm$ 0.03 & 0.01$\pm$ 0.03 & 0.48$\pm$ 0.30&I\\
\hline
\multicolumn{12}{|l|}{MKW 9}\\
\hline
1& 292& 233& 1.4$^{+ 0.1}_{-0.1}$& 2.76&   21&1$\pm$ 3.28 & 0.10$\pm$ 0.29 & 0.00$\pm$ 0.11 & 0.13$\pm$ 0.42 & 0.00$\pm$ 0.84&CI\\
2& 228& 233& 1.7$^{+ 0.2}_{-0.2}$& 3.08&   12&1$\pm$ 0.26 & 0.00$\pm$ 0.03 & 0.00$\pm$ 0.03 & 0.00$\pm$ 0.12 & 0.00$\pm$ 0.15&I\\
3& 212& 233& 2.5$^{+ 0.3}_{-0.4}$& 2.79&   36&1$\pm$ 0.22 & 0.00$\pm$ 0.55 & 0.01$\pm$ 0.06 & 0.06$\pm$ 0.06 & 0.00$\pm$ 0.08&I\\
\hline
\multicolumn{12}{|l|}{2A 0335+096}\\	
\hline
1& 492& 395& 1.6$^{+ 0.1}_{-0.1}$& 4.30&  0.7&1$\pm$ 0.11 & 0.11$\pm$ 0.02 & 0.04$\pm$ 0.02 & 0.00$\pm$ 0.03 & 0.00$\pm$ 0.05&\\
2& 412& 395& 2.1$^{+ 0.1}_{-0.1}$& 4.03&  1.8&1$\pm$ 0.08 & 0.06$\pm$ 0.03 & 0.03$\pm$ 0.01 & 0.00$\pm$ 0.03 & 0.00$\pm$ 0.02&\\
3& 406& 395& 2.8$^{+ 0.2}_{-0.1}$& 3.60&  6.0&1$\pm$ 0.12 & 0.14$\pm$ 0.12 & 0.03$\pm$ 0.01 & 0.00$\pm$ 0.01 & 0.04$\pm$ 0.05&\\
4& 461& 395& 3.8$^{+ 1.0}_{-0.8}$& 3.20&   18&1$\pm$ 0.69 & 0.62$\pm$ 0.69 & 0.00$\pm$ 0.40 & 0.02$\pm$ 0.03 & 0.02$\pm$ 0.16&CI\\
5& 418& 395& 3.5$^{+ 4.0}_{-0.5}$& 3.07&   23&1$\pm$ 0.55 & 0.12$\pm$ 1.09 & 0.03$\pm$ 0.04 & 0.02$\pm$ 0.05 & 0.00$\pm$ 0.18&CI\\
\hline
\multicolumn{12}{|l|}{MKW 3s}\\
\hline
1& 242& 233& 3.1$^{+ 0.4}_{-0.2}$& 3.92&  3.1&1$\pm$ 0.12 & 0.04$\pm$ 0.16 & 0.00$\pm$ 0.01 & 0.00$\pm$ 0.01 & 0.02$\pm$ 0.02&I\\
2& 242& 233& 5.1$^{+ 4.9}_{-1.3}$& 3.50&   11&1$\pm$ 1.00 & 1.51$\pm$ 0.77 & 0.00$\pm$ 1.42 & 0.00$\pm$ 0.02 & 0.01$\pm$ 0.04&I\\
3& 230& 233& 3.5$^{+ 0.3}_{-0.1}$& 3.36&   12&1$\pm$ 0.07 & 0.00$\pm$ 0.11 & 0.00$\pm$ 0.00 & 0.00$\pm$ 0.00 & 0.03$\pm$ 0.02&I\\
4& 248& 233& 5.4$^{+ 2.2}_{-1.0}$& 2.99&   37&1$\pm$ 0.47 & 0.57$\pm$ 0.41 & 0.11$\pm$ 0.23 & 0.00$\pm$ 0.02 & 0.06$\pm$ 0.04&I\\
\hline
\multicolumn{12}{|l|}{A 2052}\\
\hline
1& 270& 233& 2.3$^{+ 0.2}_{-0.2}$& 3.95&  2.3&1$\pm$ 0.16 & 0.42$\pm$ 0.13 & 0.05$\pm$ 0.02 & 0.00$\pm$ 0.02 & 0.00$\pm$ 0.04&\\
2& 252& 233& 2.8$^{+ 0.3}_{-0.2}$& 3.76&  4.2&1$\pm$ 0.12 & 0.15$\pm$ 0.12 & 0.01$\pm$ 0.01 & 0.00$\pm$ 0.01 & 0.01$\pm$ 0.01&\\
3& 223& 233& 3.4$^{+ 1.7}_{-0.4}$& 3.35&   12&1$\pm$ 0.49 & 0.26$\pm$ 0.50 & 0.00$\pm$ 0.01 & 0.01$\pm$ 0.01 & 0.03$\pm$ 0.03&CI\\
4& 250& 233& 4.1$^{+ 1.1}_{-0.7}$& 3.04&   28&1$\pm$ 0.53 & 0.74$\pm$ 0.55 & 0.00$\pm$ 1.21 & 0.00$\pm$ 0.01 & 0.01$\pm$ 0.04&CI\\
5& 230& 233& 3.3$^{+ 0.2}_{-0.2}$& 2.96&   29&1$\pm$ 0.07 & 0.00$\pm$ 1.04 & 0.00$\pm$ 0.02 & 0.01$\pm$ 0.02 & 0.05$\pm$ 0.07&CI\\
\hline\noalign{\smallskip}
\end{tabular}
}
\end{table*}

\begin{table*}[!ht]
\caption{Spectral fit results for the multi-temperature model (continued). 
Same notation as Table~\ref{tab:fitpar1}.}
\label{tab:fitpar2}
\centerline{
\begin{tabular}{|lrrrrr|rrrrr|r|}
\hline
\# & $\chi^2$ & d.o.f. & $kT_1$ & $\log n_{\mathrm H}$  & $\tau_{\mathrm{\hbox{cool}}}$ 
  & $Y_{1}$ &$Y_{2}$ &$Y_{3}$ &$Y_{4}$ &$Y_{5}$ & \\
   &          &        & (keV)&   (m$^{-3}$)          & (Gyr) &&&&&& \\
\hline
\multicolumn{12}{|l|}{A 4059}\\
\hline
1& 446& 395& 2.4$^{+ 0.9}_{-0.2}$& 3.75&  3.8&1$\pm$ 0.23 & 0.17$\pm$ 0.21 & 0.02$\pm$ 0.03 & 0.00$\pm$ 0.02 & 0.00$\pm$ 0.01&I\\
2& 398& 395& 5.0$^{+ 5.1}_{-1.8}$& 3.37&   15&1$\pm$ 1.21 & 1.48$\pm$ 0.89 & 0.11$\pm$ 1.09 & 0.00$\pm$ 0.25 & 0.00$\pm$ 0.03&CI\\
3& 393& 395& 4.8$^{+ 2.0}_{-0.8}$& 3.26&   19&1$\pm$ 0.47 & 0.39$\pm$ 0.46 & 0.00$\pm$ 0.05 & 0.00$\pm$ 0.01 & 0.01$\pm$ 0.02&CI\\
4& 377& 395& 5.4$^{+ 2.6}_{-1.7}$& 2.95&   41&1$\pm$ 0.93 & 0.90$\pm$ 1.03 & 0.00$\pm$ 0.86 & 0.00$\pm$ 0.02 & 0.00$\pm$ 0.10&CI\\
\hline
\multicolumn{12}{|l|}{Hydra A}\\ 
\hline
1& 283& 233& 3.6$^{+ 1.0}_{-0.6}$& 4.10&  2.3&1$\pm$ 0.44 & 0.44$\pm$ 0.45 & 0.00$\pm$ 0.03 & 0.00$\pm$ 0.02 & 0.04$\pm$ 0.08&CI\\
2& 240& 233& 3.4$^{+ 0.1}_{-0.1}$& 3.82&  4.1&1$\pm$ 0.04 & 0.00$\pm$ 0.19 & 0.00$\pm$ 0.01 & 0.000$\pm$ 0.004 & 0.07$\pm$ 0.04&I\\
3& 225& 233& 4.2$^{+ 4.1}_{-0.6}$& 3.30&   16&1$\pm$ 0.48 & 0.41$\pm$ 0.34 & 0.06$\pm$ 0.34 & 0.00$\pm$ 0.08 & 0.04$\pm$ 0.05&CI\\
4& 254& 233& 3.5$^{+ 0.6}_{-0.3}$& 3.13&   21&1$\pm$ 0.18 & 0.07$\pm$ 0.28 & 0.00$\pm$ 0.02 & 0.00$\pm$ 0.01 & 0.13$\pm$ 0.08&I\\
\hline
\multicolumn{12}{|l|}{A 496}\\
\hline
1& 245& 233& 2.4$^{+ 0.2}_{-0.2}$& 4.12&  1.6&1$\pm$ 0.12 & 0.15$\pm$ 0.08 & 0.01$\pm$ 0.01 & 0.00$\pm$ 0.03 & 0.00$\pm$ 0.02&I\\
2& 236& 233& 2.6$^{+ 0.1}_{-0.1}$& 3.85&  3.2&1$\pm$ 0.07 & 0.00$\pm$ 0.55 & 0.00$\pm$ 0.01 & 0.00$\pm$ 0.01 & 0.07$\pm$ 0.05&I\\
3& 255& 233& 4.5$^{+ 1.1}_{-0.7}$& 3.42&   12&1$\pm$ 0.40 & 0.53$\pm$ 0.42 & 0.00$\pm$ 0.60 & 0.00$\pm$ 0.02 & 0.06$\pm$ 0.03&CI\\
4& 271& 233& 4.1$^{+ 0.5}_{-0.3}$& 3.24&   18&1$\pm$ 0.09 & 0.00$\pm$ 1.09 & 0.02$\pm$ 0.04 & 0.00$\pm$ 0.02 & 0.00$\pm$ 0.03&I\\
5& 170& 233& 5.0$^{+ 7.7}_{-1.1}$& 3.00&   35&1$\pm$ 0.41 & 0.30$\pm$ 1.10 & 0.02$\pm$ 0.31 & 0.01$\pm$ 0.02 & 0.00$\pm$ 0.04&I\\
\hline
\multicolumn{12}{|l|}{A 3112}\\
\hline
1& 286& 233& 3.1$^{+ 0.1}_{-0.1}$& 4.08&  2.1&1$\pm$ 0.05 & 0.00$\pm$ 0.08 & 0.00$\pm$ 0.00 & 0.00$\pm$ 0.00 & 0.10$\pm$ 0.03&I\\
2& 268& 233& 7.5$^{+ 1.5}_{-0.9}$& 3.58&   12&1$\pm$ 0.19 & 0.00$\pm$ 0.42 & 0.87$\pm$ 0.18 & 0.00$\pm$ 0.03 & 0.00$\pm$ 0.02&C\\
3& 297& 233& 6.8$^{+ 0.8}_{-0.7}$& 3.28&   22&1$\pm$ 0.15 & 0.00$\pm$ 0.05 & 0.46$\pm$ 0.12 & 0.00$\pm$ 1.02 & 0.08$\pm$ 0.23&C\\
\hline
\multicolumn{12}{|l|}{A 1795}\\
\hline
1& 229& 233& 4.9$^{+ 1.6}_{-1.3}$& 3.99&  3.6&1$\pm$ 0.77 & 0.90$\pm$ 0.69 & 0.00$\pm$ 0.19 & 0.01$\pm$ 0.02 & 0.01$\pm$ 0.04&CI\\
2& 205& 233& 4.3$^{+ 0.4}_{-0.1}$& 3.85&  4.6&1$\pm$ 0.08 & 0.00$\pm$ 0.14 & 0.02$\pm$ 0.01 & 0.00$\pm$ 0.01 & 0.03$\pm$ 0.01&I\\
3& 207& 233& 8.0$^{+ 2.0}_{-1.2}$& 3.41&   18&1$\pm$ 0.23 & 0.01$\pm$ 0.35 & 0.38$\pm$ 0.18 & 0.00$\pm$ 0.02 & 0.00$\pm$ 0.01&\\
4& 187& 233& 6.7$^{+ 4.6}_{-1.3}$& 3.13&   31&1$\pm$ 0.33 & 0.31$\pm$ 1.01 & 0.00$\pm$ 1.32 & 0.00$\pm$ 0.02 & 0.00$\pm$ 0.02&CI\\
\hline
\multicolumn{12}{|l|}{A 399}\\
\hline
1& 285& 233& 2.6$^{+ 0.7}_{-0.5}$& 3.38&  9.3&1$\pm$ 0.40 & 0.00$\pm$ 0.28 & 0.00$\pm$ 0.10 & 0.00$\pm$ 0.07 & 0.00$\pm$ 0.16&I\\
2& 238& 233& 4.9$^{+ 1.6}_{-1.3}$& 3.30&   17&1$\pm$ 0.13 & 0.00$\pm$ 1.13 & 0.00$\pm$ 0.33 & 0.03$\pm$ 0.21 & 0.00$\pm$ 0.18&CI\\
\hline
\multicolumn{12}{|l|}{A 3266}\\
\hline
2& 299& 233& 7.7$^{+ 2.0}_{-1.3}$& 3.38&   19&1$\pm$ 0.18 & 0.00$\pm$ 1.17 & 0.00$\pm$ 0.92 & 0.00$\pm$ 0.23 & 0.00$\pm$ 0.01&CI\\
\hline
\multicolumn{12}{|l|}{Perseus}\\
\hline
2& 272& 233& 5.2$^{+ 1.0}_{-0.7}$& 4.11&  2.7&1$\pm$ 0.37 & 1.33$\pm$ 0.23 & 0.23$\pm$ 0.24 & 0.00$\pm$ 0.01 & 0.08$\pm$ 0.02&I\\
3& 474& 233& 4.2$^{+ 0.4}_{-0.3}$& 4.04&  3.0&1$\pm$ 0.15 & 0.48$\pm$ 0.15 & 0.02$\pm$ 0.02 & 0.00$\pm$ 0.01 & 0.06$\pm$ 0.01&I\\
4& 327& 233& 4.9$^{+ 0.6}_{-0.5}$& 3.80&  5.5&1$\pm$ 0.26 & 0.63$\pm$ 0.25 & 0.00$\pm$ 0.15 & 0.00$\pm$ 0.01 & 0.04$\pm$ 0.01&CI\\
5& 290& 233& 5.7$^{+ 1.2}_{-0.9}$& 3.62&  9.0&1$\pm$ 0.27 & 0.27$\pm$ 0.25 & 0.07$\pm$ 0.07 & 0.00$\pm$ 0.01 & 0.03$\pm$ 0.01&I\\
6& 195& 233& 6.4$^{+ 1.4}_{-0.6}$& 3.43&   15&1$\pm$ 0.09 & 0.00$\pm$ 0.71 & 0.11$\pm$ 0.10 & 0.00$\pm$ 0.01 & 0.03$\pm$ 0.01&I\\
\hline
\multicolumn{12}{|l|}{Coma}\\
\hline
1& 302& 233& 5.0$^{+ 1.5}_{-1.1}$& 3.29&   18&1$\pm$ 0.79 & 0.00$\pm$ 0.90 & 0.00$\pm$ 0.57 & 0.41$\pm$ 0.66 & 0.00$\pm$ 1.27&CI\\
2& 244& 233&11.2$^{+ 8.6}_{-4.2}$& 3.36&   24&1$\pm$ 0.23 & 0.00$\pm$ 1.16 & 0.00$\pm$ 1.11 & 0.00$\pm$ 0.08 & 0.00$\pm$ 0.02&CI\\
3& 226& 233& 8.0$^{+ 4.7}_{-1.9}$& 3.19&   29&1$\pm$ 0.13 & 0.00$\pm$ 1.07 & 0.00$\pm$ 0.25 & 0.00$\pm$ 0.24 & 0.14$\pm$ 0.15&CI\\
\hline
\multicolumn{12}{|l|}{A 754}\\
\hline
1& 279& 233& 7.6$^{+25.3}_{-4.0}$& 3.08&   37&1$\pm$ 1.89 & 0.00$\pm$ 3.85 & 0.16$\pm$ 1.61 & 0.00$\pm$ 3.23 & 0.00$\pm$ 0.07&CI\\
2& 227& 233& 7.0$^{+19.3}_{-1.4}$& 3.30&   21&1$\pm$ 0.18 & 0.00$\pm$ 0.91 & 0.00$\pm$ 0.43 & 0.01$\pm$ 0.41 & 0.04$\pm$ 0.12&CI\\
3& 232& 233&11.1$^{+ 4.3}_{-4.1}$& 3.08&   48&1$\pm$ 0.93 & 0.66$\pm$ 1.30 & 0.30$\pm$ 0.70 & 0.00$\pm$ 0.17 & 0.05$\pm$ 0.05&CI\\
\hline
\multicolumn{12}{|l|}{A 1835}\\
\hline
1& 150& 152& 6.2$^{+ 1.8}_{-0.6}$& 4.09&  3.2&1$\pm$ 0.27 & 0.08$\pm$ 0.45 & 0.08$\pm$ 0.09 & 0.01$\pm$ 0.01 & 0.02$\pm$ 0.02&I\\
2& 145& 152& 8.0$^{+ 0.0}_{-0.1}$& 3.57&   12&1$\pm$ 0.03 & 0.00$\pm$ 0.03 & 0.00$\pm$ 0.09 & 0.08$\pm$ 0.05 & 0.07$\pm$ 0.14&I\\
\hline\noalign{\smallskip}
\end{tabular}
}
\end{table*}

In most cases (except for the innermost annuli of 2A~0335+096 and A~2052)
two temperature components are sufficient to describe the spectrum, i.e.  in
those cases $Y_3$ is within its error bars consistent with zero.  This does not
necessarily mean that $Y_3=0$ in a strict mathematical sense, the error bars are
such that in most cases we cannot exclude the presence of emission at $T=0.25
T_1$ at a level below 1~\% of the emission measure $Y_1$ of the hottest gas.

\begin{figure}
\resizebox{\hsize}{!}{\includegraphics[angle=-90]{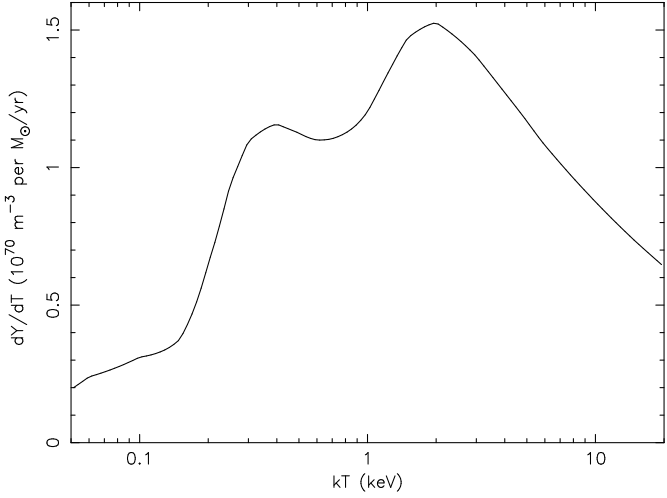}}
\caption{Differential emission measure distribution for the
isobaric cooling flow model, for 0.5 times solar abundances.}
\label{fig:cfl}
\end{figure}

We have also investigated whether these emission measure distributions are
consistent with the (cut-off) isobaric cooling flow model.  The differential
emission measure distribution ${\rm d}Y(T)/{\rm d}T$ for the isobaric cooling
flow model can be written as (see for example Fabian \cite{fabian94}):
\begin{equation}
\label{eqn:isocf}
D(T)\equiv 
{\rm d}Y(T)/{\rm d}T = {5\dot{M}k \over 2\mu m_{\rm H} \Lambda(T),}
\end{equation}
where $\dot{M}$ is the mass deposition rate, $k$ is Boltzmann's constant, $\mu$
the mean molecular weight (0.618 for a plasma with 0.5 times solar abundances),
$m_{\rm H}$ is the mass of a hydrogen atom, and $\Lambda(T)$ is the cooling
function.  The function $D(T)$ for a 0.5 times solar abundances plasma is shown
in Fig.~\ref{fig:cfl}.  For any given cluster of galaxies and shell, the
function $D(T)$ should be put to zero naturally above a temperature $T_{\max}$,
corresponding to the hottest gas at that location.  It is seen from
Fig.~\ref{fig:cfl} that for the relevant part of $D(T)$ for most of our data
(between 0.3--3~keV) this function does not vary by more than 20~\% around a
mean value of $1.3\times 10^{70}$~m$^{-3}$M$_{\sun}^{-1}$yr.  Hence, since in
our modeling we have taken temperature steps of a factor of 2, the emission
measures $Y_i$ for each temperature component $T_i$ should scale approximately
as $Y_i\sim T_i$.  In several cases the deduced emission measure distribution
decreases more steeply towards lower temperatures than proportional to $T$
(Fig.~\ref{fig:cf_a262}), or may be even cut-off.  Therefore we modeled the
emission measures of Table~\ref{tab:fitpar1}--\ref{tab:fitpar2} with the
function $D(T)$, but cut-off to zero outside the interval ($T_{\min},T_{\max}$),
where $T_{\min}$ is a cut-off temperature.  For the standard isobaric cooling
flow model, $T_{\min}$ is zero.

We cannot assume that the division boundaries between the temperature components
are exactly spaced by factors of two in temperature, for an arbitrary emission
measure model.  The division boundaries should be chosen in such a way that the
emission measure weighted temperature for each bin equals the temperature
assigned to it in our spectral fitting procedure.  We adopted an iterative
procedure where for a given set of parameters ($T_{\min},T_{\max}$) and the
emission measure distribution $D(T)$ the bin boundaries were adjusted until the
correct weighted average temperatures were determined.  Then in a least squares
fit the normalisation $\dot{M}$ was determined.

In our grid search we have put an artificial lower limit of 0.1~keV for
$T_{\min}$; XMM-Newton is not sensitive to lower temperatures.

In a number of cases we find that the spectrum is consistent with
$T_{\min}=T_{\max}$, i.e.  the spectrum is formally indistinguishable from an
isothermal plasma.  This is the case for truly isothermal plasmas or spectra
with poorer statistics, both conditions occurring in general in the outer parts
of clusters.  We indicate those cases with the letter $I$ in the last column of
Table~\ref{tab:fitpar1}--\ref{tab:fitpar2}.

In other cases we find that $T_{\min}$ is less than 0.1~keV, hence the spectrum
is formally consistent with the standard (non-cut off) isobaric cooling flow
model, as far as the sensitivity range of XMM-Newton is concerned.  Again, this
is the case for truly isobaric cooling plasmas, or for spectra with poorer
statistics.  We indicate those cases with the letter $C$ in the last column of
Table~\ref{tab:fitpar1}--\ref{tab:fitpar2}.

There are several instances where the statistics do not allow us to distinguish
the full isobaric cooling flow model (C) from the purely isothermal model (I).
These are the shells for which we list both a C and I in the last column of
Table~\ref{tab:fitpar1}.  In all other cases we have either a good lower limit
to $T_{\min}$ or upper limit to $T_{\max}$.

Finally, in a few rare cases (Virgo shells 6--8, Perseus shells 2--6) the
emission measure for $T_5$ has formally a significant non-zero value, despite
the fact that the table indicates that the spectrum is consistent with an
isothermal model ("I").  This interpretation is chosen due to the fact that
$Y_2$--$Y_4$ are zero within the error bars, and therefore the best fit to the
isobaric cooling flow model finds $T_{\min}=T_{\max}$.  Note however that the
flux associated with $T_5$ is very small, for example $<0.4$~\% of the
0.2--10~keV flux of shell 6 for Virgo, due to both the relatively small emission
measure and the low temperature of $\sim$0.2~keV.  These emission measures are
well within the systematic uncertainties and we are only able to "detect" them
thanks to the high statistical quality of the Virgo spectrum.  In yet another
case (S\'ersic~159$-$3) the enhancement of $Y_5$ for shells 2--4 despite the
negligible values for $Y_3$ and $Y_4$ is due to the strong soft X-ray excess in
this cluster (see Kaastra et al.  \cite{kaastra03}).

\begin{table*}[!ht]
\caption{Constraints for the isobaric cooling flow model.  The lower limit,
best-fit value and upper limit are given for both $T_{\min}$ (in keV) and
$T_{\min}/T_{\max}$.  Only shells with a cooling time less than 30~Gyr (as
determined from the single temperature fits) are listed.  No entries are listed
whenever the data cannot constrain the ratio $T_{\min}/T_{\max}$ (i.e., any
value between 0--1 is statistically allowed).  The \#\ indicates the shell
number.}
\label{tab:fittemp}
\centerline{
\begin{tabular}{|l|rrr|rrr|rrr|rrr|rrr|rrr|}
\hline
\multicolumn{1}{|c|}{\#} & 
\multicolumn{3}{|c|}{$T_{\min}$} &
\multicolumn{3}{|c|}{$T_{\min}/T_{\max}$} & 
\multicolumn{3}{|c|}{$T_{\min}$} &
\multicolumn{3}{|c|}{$T_{\min}/T_{\max}$} & 
\multicolumn{3}{|c|}{$T_{\min}$} &
\multicolumn{3}{|c|}{$T_{\min}/T_{\max}$} \\
  & 
  low & best & upp & low & best & upp & 
  low & best & upp & low & best & upp & 
  low & best & upp & low & best & upp \\
\hline
\multicolumn{7}{|l|}{NGC 533}&
\multicolumn{6}{|l|}{Virgo}&
\multicolumn{6}{|l|}{A 262}\\
\hline
1&   0.2& 0.7& 1.3&   0.21&0.94&1 &    &    &    &     &    &     &   0.5& 0.6& 1.3&   0.23&0.32&0.92 \\
2&$<$0.1& 0.6& 2.1&      0&0.45&1 & 0.8& 1.5& 1.6& 0.31&0.45&0.78 &   0.7& 1.1& 1.5&   0.32&0.47&0.86 \\
3&   0.4& 0.7& 1.8&   0.15&0.25&1 & 0.8& 1.4& 1.6& 0.33&0.46&0.72 &   0.6& 1.4& 4.0&   0.20&0.46&1    \\
4&   0.4& 0.6& 2.7&   0.18&0.29&1 & 0.9& 1.5& 1.9& 0.32&0.47&0.86 &   1.0& 2.0& 4.3&   0.30&0.48&1    \\
5&   0.4& 0.7& 2.9&   0.20&0.31&1 & 1.0& 1.7& 2.0& 0.32&0.47&0.90 &$<$0.1& 1.4& 6.5&      0&0.33&1    \\
6&      &    &    &       &    &  & 0.7& 1.6& 4.4& 0.22&0.44&1    &   0.9& 1.6& 4.5&   0.27&0.48&1    \\
7&      &    &    &       &    &  & 0.8& 1.3& 4.9& 0.22&0.34&1    &      &    &    &       &    &     \\
8&      &    &    &       &    &  & 0.8& 1.1& 5.7& 0.20&0.27&1    &      &    &    &       &    &     \\
\hline
\multicolumn{7}{|l|}{A 1837}&
\multicolumn{6}{|l|}{S\'ersic~159$-$3}&
\multicolumn{6}{|l|}{MKW 9}\\
\hline
1&   1.1& 1.5& 5.9& 0.25  &0.31&1    & 1.1& 1.1& 2.3& 0.31&0.33&0.91&$<$0.1& 1.1& 2.7&    0  &0.44&1 \\
2&$<$0.1& 1.8& 3.3&    0  &0.16&0.29 & 1.0& 2.5& 2.8& 0.25&0.80&1   &   0.8& 1.6& 3.2& 0.32  &0.94&1 \\
3&      &    &    &       &    &     & 0.9& 2.6& 4.8& 0.24&0.94&1   &   0.3& 1.2& 4.7& 0.08  &0.31&1 \\
4&      &    &    &       &    &     & 0.9& 2.6& 5.1& 0.25&0.93&1   &      &    &    &       &    &  \\
\hline
\multicolumn{7}{|l|}{2A 0335+096}&
\multicolumn{6}{|l|}{MKW 3s}&
\multicolumn{6}{|l|}{A 2052}\\
\hline
1& 0.6& 0.7& 0.7& 0.26&0.28&0.30& 1.3& 1.6& 5.9& 0.28&0.35&1 & 0.5& 0.6& 1.1& 0.17&0.21&0.26 \\
2& 0.9& 0.9& 1.0& 0.29&0.30&0.31& 0.6& 1.3& 4.8& 0.09&0.19&1 & 1.1& 1.2& 1.4& 0.26&0.29&0.32 \\
3& 1.0& 1.1& 1.3& 0.25&0.28&0.31& 1.5& 2.5& 6.5& 0.30&0.56&1 &    &    &    &     &    &     \\
4&    &    &    &     &    &    & 0.7& 1.5& 5.2& 0.09&0.21&1 &    &    &    &     &    &     \\
\hline
\multicolumn{7}{|l|}{A 4059}&
\multicolumn{6}{|l|}{Hydra A (A 780)}&
\multicolumn{6}{|l|}{A 496}\\
\hline
1& 0.7& 1.0& 4.6& 0.21&0.27&1 &$<$0.1& 2.9& 6.8&      0&0.72&1 &   0.9& 1.0& 2.3&   0.27&0.29&0.93 \\
2&    &    &    &     &    &  &   1.4& 3.7& 6.3&   0.28&0.91&1 &   0.7& 2.5& 4.8&   0.20&0.94&1    \\
3&    &    &    &     &    &  &$<$0.1& 1.4& 7.9&      0&0.24&1 &$<$0.1& 2.6& 8.4&      0&0.29&1    \\ 
4&    &    &    &     &    &  &   1.2& 2.8& 6.7&   0.25&0.47&1 &   0.9& 1.9& 7.6&   0.18&0.30&1    \\ 
5&    &    &    &     &    &  &      &    &    &       &    &  &   0.4& 1.9& 9.2&   0.05&0.27&1    \\  
\hline
\multicolumn{7}{|l|}{A 3112}&
\multicolumn{6}{|l|}{A 1795}&
\multicolumn{6}{|l|}{A 399}\\
\hline
1&  1.4 & 2.9 & 5.7 &   0.31& 0.88 & 1    &$<$0.1& 3.9& 9.2&      0&0.72&1   & 0.6 & 2.6 & 4.8 & 0.17 & 1 & 1 \\
2&$<$0.1& 1.0 & 2.1 &      0& 0.09 & 0.19 &   1.8& 2.1& 8.2&   0.28&0.31&1   &     &     &     &      &   &   \\ 
3&$<$0.1& 1.6 & 2.1 &      0& 0.16 & 0.19 &   1.0& 2.0& 3.1&   0.09&0.16&0.24&     &     &     &      &   &   \\ 
\hline
\multicolumn{7}{|l|}{Perseus (A 426)}&
\multicolumn{6}{|l|}{A 1835}&
\multicolumn{6}{|l|}{ }\\
\hline
1&      &      &    &     &     &  & 1.5& 2.6& 11.6&0.18&0.27&1 &      &      &      &      &   &   \\
2&   1.0&  1.8 & 4.9& 0.15& 0.29& 1& 3.3& 7.8& 14.8&0.26&0.93&1 &      &      &      &      &   &   \\
3&   1.2&  1.4 & 4.0& 0.22& 0.24& 1&    &    &     &    &    &  &      &      &      &      &   &   \\
4&$<$0.1&  2.0 & 4.7&    0& 0.25& 1&    &    &     &    &    &  &      &      &      &      &   &   \\
5&   1.5&  2.1 & 5.6& 0.19& 0.25& 1&    &    &     &    &    &  &      &      &      &      &   &   \\
6&   0.5&  2.6 &12.2& 0.06& 0.26& 1&    &    &     &    &    &  &      &      &      &      &   &   \\
\hline\noalign{\smallskip}
\end{tabular}
}
\end{table*}

\subsection{Alternative spectral models\label{sect:wdem}}

The observed differential emission measure distribution in several of our
clusters shows a rather steep temperature gradient downwards from the maximum
temperature, as our multi-temperature fitting shows.  Here we try to
parameterise this behaviour, by fitting the spectra with a DEM model of the
following shape:
\begin{equation}
\label{eqn:demalpha}
{ {\mathrm d}Y \over {\mathrm d}T } = \left\{
\begin{array}{ll}
c T^{\alpha} &  \qquad\mbox{if $T < T_{\max}$};\\
 0 & \qquad\mbox{if $T \geq T_{\max}$}.
 \end{array} \right.
\end{equation}

\begin{table*}[!ht]
\caption{Spectral fit results with the {\sl wdem} model.}
\label{tab:fitalpha}
\centerline{
\begin{tabular}{|l|rrlr|rrlr|rrlr|}
\hline
\# & 
 $\chi^2$ & $T_{\max}$ & $1/\alpha$ & $\log n_{\mathrm H}$ &
 $\chi^2$ & $T_{\max}$ & $1/\alpha$ & $\log n_{\mathrm H}$ &
 $\chi^2$ & $T_{\max}$ & $1/\alpha$ & $\log n_{\mathrm H}$ \\
   & 
            & (keV)      &            & (m$^{-3}$) &
            & (keV)      &            & (m$^{-3}$) &
            & (keV)      &            & (m$^{-3}$) \\
\hline
\multicolumn{1}{|l|}{ }&
\multicolumn{4}{|l|}{NGC 533} &
\multicolumn{4}{|l|}{Virgo} &
\multicolumn{4}{|l|}{A 262} \\
\hline
1& 378&   1.04$\pm$0.05& 0.86$^{+0.34}_{-0.21}$& 3.71 &    &                &              &      & 325&   1.72$\pm$0.10& 0.78$\pm$0.15& 3.98 \\
2& 286&   1.41$\pm$0.10& 0.50$\pm$0.12         & 3.13 & 366&   1.99$\pm$0.05& 0.31$\pm$0.02& 4.45 & 215&   1.93$\pm$0.14& 0.28$\pm$0.06& 3.67 \\
3& 210&   1.94$\pm$0.22& 0.48$\pm$0.13         & 2.80 & 327&   1.93$\pm$0.04& 0.25$\pm$0.02& 4.12 & 241&   2.46$\pm$0.17& 0.29$\pm$0.07& 3.36 \\
4& 271&   1.74$\pm$0.46& 0.26$^{+0.29}_{-0.26}$& 2.64 & 275&   2.29$\pm$0.06& 0.25$\pm$0.03& 3.96 & 232&   2.57$\pm$0.29& 0.23$^{+0.12}_{-0.23}$& 3.21 \\
5& 264&   1.83$\pm$0.69& 0.39$^{+0.57}_{-0.30}$& 2.51 & 239&   2.33$\pm$0.08& 0.24$\pm$0.03& 3.82 & 257&   2.98$\pm$0.36& 0.43$\pm$0.18& 3.06 \\
6& 237&   1.71$\pm$0.42& 0.42$^{+0.51}_{-0.19}$& 2.30 & 292&   2.59$\pm$0.09& 0.29$\pm$0.04& 3.65 & 267&   2.58$\pm$0.40& 0.21$^{+0.15}_{-0.21}$& 2.93 \\
7&    &                &                       &      & 276&   2.82$\pm$0.13& 0.23$\pm$0.07& 3.50 & 244&   2.26$\pm$0.42& 0.00$^{+0.34}_{-0.00}$& 2.71 \\ 
8&    &                &                       &      & 253&   3.05$\pm$0.17& 0.25$\pm$0.08& 3.37 & 212&   2.42$\pm$0.50& 0.18$\pm$0.20& 2.55 \\
\hline
\multicolumn{1}{|l|}{ }&
\multicolumn{4}{|l|}{A 1837} &
\multicolumn{4}{|l|}{S\'ersic~159$-$3} &
\multicolumn{4}{|l|}{MKW 9} \\
\hline
1& 231&   3.82$\pm$0.79& 0.34$^{+0.19}_{-0.34}$& 3.46 & 281&   2.59$\pm$0.12& 0.22$\pm$0.06         & 4.12 & 299&   1.50$\pm$0.26& 0.17$\pm$0.17         & 3.28 \\
2&    &                &                       &      & 283&   2.80$\pm$0.16& 0.21$\pm$0.08         & 3.77 & 228&   1.67$\pm$0.33& 0.00$^{+0.20}_{-0.00}$& 3.08 \\
3&    &                &                       &      & 277&   2.76$\pm$0.26& 0.16$^{+0.07}_{-0.16}$& 3.35 & 216&   3.49$\pm$0.85& 0.67$^{+0.70}_{-0.40}$& 2.81 \\
4&    &                &                       &      & 275&   2.79$\pm$0.45& 0.19$^{+0.25}_{-0.19}$& 2.98 & 239&   2.53$\pm$0.93& 0.04$^{+0.53}_{-0.04}$& 2.58 \\
5&    &                &                       &      & 269&   3.11$\pm$0.67& 0.34$^{+0.38}_{-0.34}$& 2.73 &    &                &                       &      \\
\hline
\multicolumn{1}{|l|}{ }&
\multicolumn{4}{|l|}{2A 0335+096} &
\multicolumn{4}{|l|}{MKW 3s} &
\multicolumn{4}{|l|}{A 2052} \\
\hline
1& 502&   2.06$\pm$0.09& 0.48$\pm$0.08         & 4.33 & 244&   3.69$\pm$0.56& 0.27$^{+0.13}_{-0.27}$& 3.93 & 276&   2.74$\pm$0.15& 0.76$\pm$0.11         & 4.03 \\
2& 419&   2.59$\pm$0.11& 0.42$\pm$0.06         & 4.05 & 246&   4.34$\pm$0.38& 0.36$^{+0.13}_{-0.17}$& 3.70 & 252&   3.31$\pm$0.14& 0.37$\pm$0.06         & 3.79 \\
3& 412&   3.46$\pm$0.17& 0.51$\pm$0.10         & 3.63 & 238&   4.00$\pm$0.54& 0.20$^{+0.15}_{-0.20}$& 3.37 & 231&   3.62$\pm$0.35& 0.28$^{+0.11}_{-0.18}$& 3.40 \\
4& 464&   3.92$\pm$0.44& 0.46$\pm$0.22         & 3.30 & 255&   5.83$\pm$0.58& 0.81$^{+0.43}_{-0.27}$& 3.11 & 256&   3.75$\pm$0.52& 0.25$^{+0.13}_{-0.25}$& 3.16 \\
5& 419&   4.70$\pm$0.80& 0.79$^{+0.81}_{-0.39}$& 3.10 & 257&   3.53$\pm$0.66& 0.03$^{+0.37}_{-0.03}$& 2.89 & 234&   4.15$\pm$0.86& 0.39$^{+0.28}_{-0.39}$& 2.97 \\
6&    &                &                       &      &    &                &                       &      & 239&   3.88$\pm$0.49& 0.31$\pm$0.21         & 2.77 \\
\hline
\multicolumn{1}{|l|}{ }&
\multicolumn{4}{|l|}{A 4059} &
\multicolumn{4}{|l|}{Hydra A (A 780)} &
\multicolumn{4}{|l|}{A 496} \\
\hline
1& 446&   2.93$\pm$0.27& 0.46$\pm$0.11         & 3.79 & 286&   3.68$\pm$0.56& 0.31$^{+0.16}_{-0.31}$& 4.18 & 247&   2.87$\pm$0.20& 0.41$\pm$0.08         & 4.15 \\
2& 399&   4.09$\pm$0.46& 0.36$\pm$0.17         & 3.58 & 251&   4.16$\pm$0.67& 0.30$^{+0.20}_{-0.30}$& 3.83 & 244&   2.86$\pm$0.39& 0.15$\pm$0.15         & 3.86 \\
3& 397&   4.94$\pm$0.81& 0.30$^{+0.17}_{-0.30}$& 3.33 & 230&   4.70$\pm$0.42& 0.64$^{+0.26}_{-0.19}$& 3.38 & 268&   4.96$\pm$0.47& 0.63$\pm$0.22         & 3.52 \\
4& 378&   4.63$\pm$0.86& 0.21$^{+0.27}_{-0.21}$& 3.09 & 262&   4.48$\pm$0.89& 0.47$^{+0.69}_{-0.34}$& 3.16 & 271&   5.43$\pm$1.45& 0.47$\pm$0.47         & 3.25 \\
5&    &                &                       &      &    &                &                       &      & 171&   5.93$\pm$1.68& 0.60$\pm$0.60         & 3.06 \\
\hline
\multicolumn{1}{|l|}{ }&
\multicolumn{4}{|l|}{A 3112} &
\multicolumn{4}{|l|}{A 1795} &
\multicolumn{4}{|l|}{A 399} \\
\hline
1& 313&   3.70$\pm$0.25& 0.27$\pm$0.09         & 4.10 & 235&   4.89$\pm$0.33& 0.53$\pm$0.14         & 4.13 & 285&   2.60$\pm$1.24& 0.00$^{+0.65}_{-0.00}$& 3.38 \\
2& 278&   5.60$\pm$0.56& 0.53$\pm$0.21         & 3.69 & 216&   5.85$\pm$0.41& 0.52$\pm$0.15         & 3.86 &    &                &                       &      \\
3&    &                &                       &      & 213&   8.20$\pm$0.75& 0.76$^{+0.29}_{-0.20}$& 3.48 &    &                &                       &      \\
4&    &                &                       &      & 188&   7.10$\pm$1.36& 0.31$^{+0.34}_{-0.31}$& 3.19 &    &                &                       &      \\
\hline
\multicolumn{1}{|l|}{ }&
\multicolumn{4}{|l|}{Perseus (A 426)} &
\multicolumn{4}{|l|}{A 1835} &
\multicolumn{4}{|l|}{ }\\
\hline
1&    &                &                       &      & 153&   9.11$\pm$0.80& 1.34$^{+0.64}_{-0.35}$& 4.13 &    &                &                       &      \\  
2& 308&   4.59$\pm$0.17& 0.68$\pm$0.09         & 4.32 &    &                &                       &      &    &                &                       &      \\
3& 528&   4.72$\pm$0.14& 0.64$\pm$0.07         & 4.11 &    &                &                       &      &    &                &                       &      \\
4& 353&   5.22$\pm$0.21& 0.58$\pm$0.09         & 3.90 &    &                &                       &      &    &                &                       &      \\
5& 301&   7.09$\pm$0.49& 0.85$\pm$0.19         & 3.68 &    &                &                       &      &    &                &                       &      \\
6& 210&   9.87$\pm$0.94& 1.69$^{+0.83}_{-0.45}$& 3.46 &    &                &                       &      &    &                &                       &      \\
7& 238&   9.46$\pm$1.35& 1.08$^{+0.60}_{-0.36}$& 3.19 &    &                &                       &      &    &                &                       &      \\
\hline\noalign{\smallskip}
\end{tabular}
}
\end{table*}

For $\alpha\rightarrow\infty$, we obtain the isothermal model, for large
$\alpha$ a steep temperature decline is recovered while for $\alpha=0$ the
emission measure distribution is flat.  Note that Peterson et al.
(\cite{peterson03}) use a similar parameterisation but then for the differential
luminosity distribution).  In practice, we have implemented the model
(\ref{eqn:demalpha}) by using the integrated emission measure $Y_{\mathrm{tot}}$
instead of $c$ for the normalisation, and instead of $\alpha$ its inverse
$p=1/\alpha$, so that we can test isothermality by taking $p=0$.  The emission
measure distribution for the model is binned to bins with logarithmic steps of
0.10 in $\log T$, and for each bin the spectrum is evaluated at the
emission measure averaged temperature and with the integrated emission measure
for the relevant bin (this is needed since for large $\alpha$ the emission
measure weighted temperature is very close to the upper temperature limit of the
bin, and not to the bin centroid).  This model is now implemented as the {\sl
wdem} model in the SPEX package.  The results of our spectral fits are displayed
in Table~\ref{tab:fitalpha}.

While multi-temperature fitting (Sect.~\ref{sect:multit}) has the advantage of
being completely model independent, the present {\sl wdem} model has the
advantage that the estimated uncertainties on the derived parameters, in
particular $T_{\max}$ and $\alpha$ (and hence the degree of non-isothermality)
are much smaller than the equivalent parameters for the multi-temperature model,
provided of course that the {\sl wdem} model applies.  This is due to the fact
that if the spectrum shows weak emission above $T_{\max}$ (either true or due to
noise), the multi-temperature model may increase $T_{\max}$ to accommodate for
this high temperature tail by adjusting all values of the emission measures
accordingly, yielding almost the same best-fit $\chi^2$, but with larger error
bars on the derived emission measures.  A good example is the second shell of
MKW~3s, where the multi-temperature model gives $\chi^2=242$ for 233 degrees of
freedom, with $T_{\max}$ in the range between 3.8--10~keV , and with a 100~\%
error on the emission measure of the hottest component.  The {\sl wdem} model
has $\chi^2=244$ for 237 degrees of freedom (i.e.  almost the same goodness of
fit), but $T_{\max}$ is now well constrained:  4.34$\pm$0.38~keV, and with
$1/\alpha=0.36^{+0.13}_{-0.17}$, the spectrum is non-isothermal at the 2$\sigma$
confidence level.  Therefore the {\sl wdem} model offers a more sensitive test
for isothermality than the multi-temperature fitting, and this is why in several
cases the {\sl wdem} model is able to detect the emission measure of the cooler
gas and correspondingly finds a significant $1/\alpha>0$ (i.e.,
non-isothermality), despite the fact that the multi-temperature fit is
consistent with an isothermal model, as indicated by an "I" in
Tables~\ref{tab:fitpar1}--\ref{tab:fitpar2}.  Note however that the emission
measure distribution from the multi-temperature fits in the cases indicated
above is not only consistent with an isothermal spectrum but in general also
with a broad range of non-isothermal models, including the solution from the
{\sl wdem} model.

\begin{figure}
\resizebox{\hsize}{!}{\includegraphics[angle=-90]{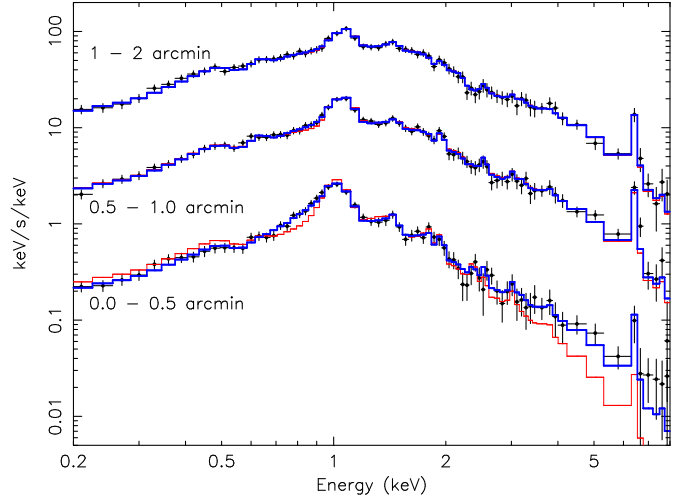}}
\caption{Spectrum of A~2052 in the inner three shells (with cooling gas).  The
spectra of all three EPIC camera's have been added.  The spectra of the
0.5--1.0\arcmin\ and 1--2\arcmin\ shells have been multiplied by factors of 5
and 25, respectively.  The spectra are shown as energy times counts/s/keV.  Thin
solid histograms:  fits with a single temperature model; thick solid histograms:
fits with the {\sl wdem} model described in the text.}
\label{fig:a2052_spec}
\end{figure}

Fig.~\ref{fig:a2052_spec} shows an example of the fit improvement with the {\sl
wdem} model as compared to a single temperature model (the best fit model for
the multi temperature model of Sect.~\ref{sect:multit} is very similar to the
fit with the {\sl wdem} model).  The single temperature model in the innermost
shells fails in two ways.  First, it under-estimates the Bremsstrahlung
continuum at high energies.  This is due to the fact that the statistical
quality of the spectrum near the Fe-L complex is much better than at high
energies, due to a larger number of photons and a higher effective area.
Therefore, the temperature of the hottest gas is dominated by the line emission
from the most highly ionised ions in the Fe-L complex.  Secondly, the single
temperature model lacks the emission from the lowly ionised Fe-L ions (with
predominantly line emission in the 0.7--0.9~keV range).  This emission is
included both in the multi-temperature fitting and in the {\sl wdem} model.  For
larger radii, the spectrum becomes close to isothermal and there all three
spectral models agree (see the 1--2\arcmin\ shell in Fig.~\ref{fig:a2052_spec}).

\begin{figure}
\resizebox{\hsize}{!}{\includegraphics[angle=-90]{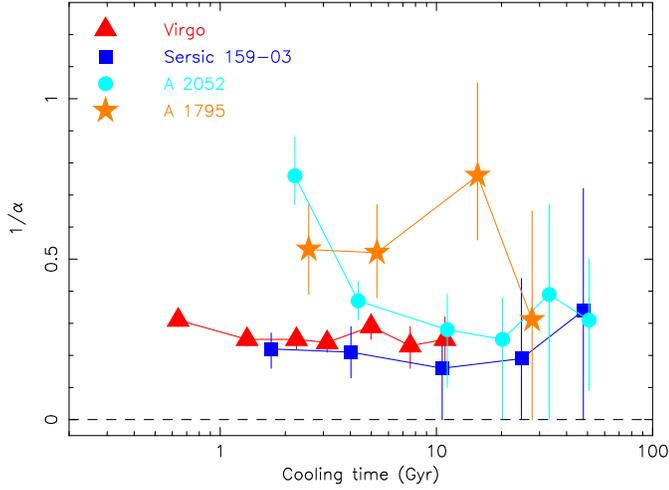}}
\caption{Steepness of the differential emission measure distribution
versus cooling time for four clusters as indicated on the plot.}
\label{fig:overalfa}
\end{figure}

In Fig.~\ref{fig:overalfa} we plot $1/\alpha$ versus cooling time for some of
our clusters.  For cooling times larger than 10--20~Gyr, the error bars on
$1/\alpha$ are in general large, and include the value 0, which indicates that
the spectra are consistent with an isothermal model in the outer parts of the
cluster.  The slightly enhanced values of $1/\alpha$ in a few rare cases at
these locations are probably due to azimuthal temperature variations or other
irregularities at larger distance from the core of the cluster.  In the cooling
region (cooling times less than 10--20~Gyr), none of our clusters shows a
significant variation of $\alpha$ versus cooling time (and hence radius).  Only
in three cases (NGC~533, A~262 and A~2052) the innermost shell shows an enhanced
value of $1/\alpha$ (see Fig.~\ref{fig:overalfa} for the case of A~2052).
Perhaps in these cases an additional soft spectral component related to the
central galaxy is present.

\begin{table}[!ht]
\caption{Weighted average temperature $T_{\max}$ of the hot gas
in the region with cooling time between 10--100~Gyr, and weighted average
value for $1/\alpha$ for the full cluster (average is dominated by the cooling
cluster core).
}
\label{tab:averalfa}
\centerline{
\begin{tabular}{|lrr|}
\hline
Cluster          & $T_{\max}$ (keV) & $1/\alpha$ \\
\hline
NGC 533          &  1.82 $\pm$0.23 &0.46$\pm$ 0.13\\
Virgo	         &  3.05 $\pm$0.24 &0.27$\pm$ 0.02\\
A 262	         &  2.59 $\pm$0.24 &0.27$\pm$ 0.06\\
A 1837           &  4.51 $\pm$0.65 &0.29$\pm$ 0.22\\
S\'ersic 159$-$3 &  3.01 $\pm$0.28 &0.20$\pm$ 0.05\\
MKW 9	         &  1.29 $\pm$0.26 &0.11$\pm$ 0.17\\
2A 0335+096      &  4.26 $\pm$0.47 &0.46$\pm$ 0.06\\
MKW 3s           &  4.55 $\pm$0.46 &0.29$\pm$ 0.11\\
A 2052           &  3.75 $\pm$0.34 &0.34$\pm$ 0.07\\
A 4059           &  4.92 $\pm$0.68 &0.36$\pm$ 0.11\\
Hydra A          &  4.68 $\pm$0.52 &0.37$\pm$ 0.16\\
A 496	         &  5.11 $\pm$0.61 &0.37$\pm$ 0.10\\
A 3112           &  9.32 $\pm$0.51 &0.30$\pm$ 0.11\\
A 1795           &  7.98 $\pm$0.92 &0.53$\pm$ 0.13\\
A 399	         &  10.8 $\pm$3.81 &0.00$\pm$ 0.65\\
A 1835           &  16.5 $\pm$1.86 &1.34$\pm$ 0.50\\
A 2199           &  5.00 $\pm$0.60 &0.38$\pm$ 0.09\\
Perseus          &  9.79 $\pm$1.04 &0.65$\pm$ 0.06\\
\hline\noalign{\smallskip}
\end{tabular}
}
\end{table}

\begin{figure}
\resizebox{\hsize}{!}{\includegraphics[angle=-90]{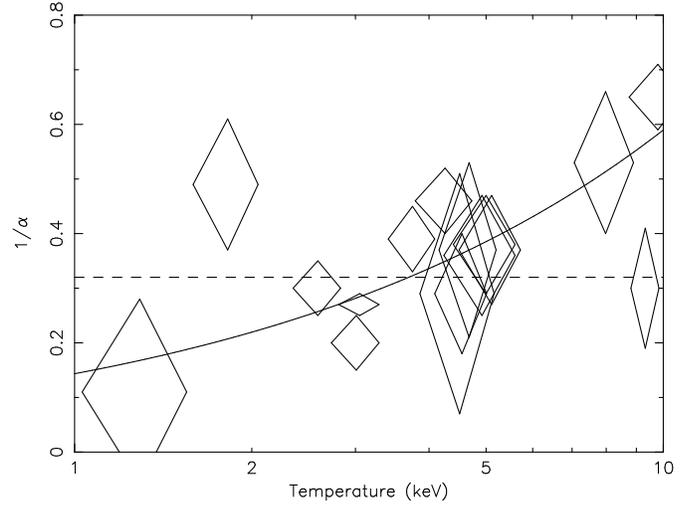}}
\caption{Steepness of the differential emission measure distribution
versus maximum temperature. 
The diamonds represent the data points with error bars.
The dashed line indicates the weighted average value of $1/\alpha$, the
solid line the best fit power law scaling as described in the text.}
\label{fig:atplot}
\end{figure}

The average value for $1/\alpha$ for our clusters is listed in
Table~\ref{tab:averalfa}.  For the three clusters mentioned above we omitted the
central shell from the average.  We also list the temperature of the hot gas as
determined in the region where the cooling time is between 10--100~Gyr.  These
temperatures are in general somewhat higher than the temperatures derived from
the single temperature fits in the same region.  This is due to the fact that in
the {\sl wdem} model used here, a certain range of temperatures is allowed and
hence the maximum temperature is higher than the average temperature.
Fig.~\ref{fig:atplot} shows the data graphically.

The weighted average value for $1/\alpha$ for all clusters is 0.32$\pm$0.02,
implying $\alpha=3.1\pm 0.2$.  Thus, at each radius in the cooling region the
emission measure distribution decreases rapidly, proportional to $T^{3.1}$.
There is a weak indication that $1/\alpha$ is slightly larger for the hotter
clusters, see Fig.~\ref{fig:atplot}.  Fitting formally a power law yields
$\alpha=(7.1\pm 1.2) T^{-0.61\pm 0.13}$, where $T$ is in keV. However, this
fit is driven by only a few cool or hot clusters, so more data is needed
to confirm such a correlation.

\section{Discussion\label{sect:discussion}}

\subsection{Present results}

The models with a single temperature at each radius (Table~\ref{tab:fit1t}) do
not always provide the best spectral fits to the inner cooling parts in our
clusters, although in some cases the single temperature model gives a $\chi^2$
value that is formally acceptable.  However, in almost all cases, the fits with
the multi-temperature model (Tables~\ref{tab:fitpar1}--\ref{tab:fitpar2}) or
{\sl wdem} model (Table~\ref{tab:fitalpha}) have much lower $\chi^2$ values, and
often show a significant deviation from isothermality (as expressed for example
in the ratio $Y_2/Y_1$ and $\alpha$, respectively), although the emission
measure distribution is dominated by emission from the hottest gas at each
radius.  As a typical example, the second shell of A~262 has $\chi^2=241$ for
237 degrees of freedom when fitted with the single temperature model, which is
formally a good fit.  But the multi-temperature model decreases $\chi^2$ to 220
(for 233 degrees of freedom), and gives a 3$\sigma$ detection of a cooler
component ($Y_2/Y_1=0.06\pm 0.02$).

However the single temperature fits show that gas cooling occurs in the center
of these clusters.  In all clusters where the cooling time at any radius is less
than 10--15~Gyr we see a statistical significant temperature drop within that
radius.  Consistent with this, the non-cooling clusters A~3266, Coma and A~754
(cooling time always larger than 20 Gyr) show no central temperature drop.

\begin{figure}
\resizebox{\hsize}{!}{\includegraphics[angle=-90]{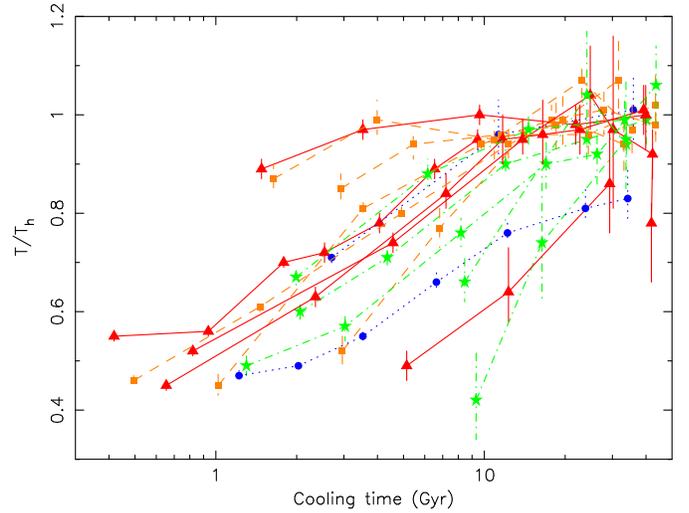}}
\caption{Scaled temperature profile versus cooling time. Temperatures
have been normalised to the temperature $T_h$ of the hot gas outside the cooling
region. Solid lines, triangles: clusters cooler than 3~keV; dashed lines,
squares: clusters with temperatures from 3--4~keV; dash-dotted lines, stars:
clusters with temperatures from 4--6~keV; dotted lines, circles:
clusters with temperatures above 6~keV. The non-cooling clusters Coma, A754
and A~3266 have been omitted from the plot.}
\label{fig:ttc}
\end{figure}

\begin{figure}
\resizebox{\hsize}{!}{\includegraphics[angle=-90]{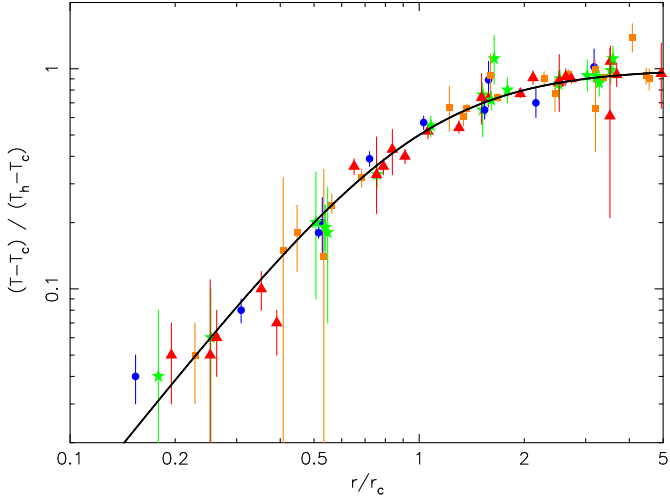}}
\caption{Scaled temperature profile versus scaled radius. Temperatures
have been normalised as indicated using
the temperature $T_h$ of the hot gas outside the cooling
region as well as the central temperature $T_c$.
Triangles: clusters cooler than 3~keV; 
squares: clusters with temperatures from 3--4~keV; stars:
clusters with temperatures from 4--6~keV; circles:
clusters with temperatures above 6~keV. The non-cooling clusters Coma, A754
and A~3266 have been omitted from the plot.
The solid line is the scaling curve Eq.~(\ref{eqn:tfit}) with $\mu=2$.}
\label{fig:trc}
\end{figure}

However, there is no universal scaling law with cooling time (or gas density).
This is illustrated in Fig.~\ref{fig:ttc}.  Despite short cooling times, in
S\'ersic~159$-$3 and MKW~3s there is only a very weak temperature gradient.
Contrary to this, in A~496 the temperature gradient is very steep. 

\begin{table}[!ht]
\caption{Characteristic temperatures and cooling times. $T_h$ is the
temperature outside the cooling region, $T_c$ the temperature in the
center of the cluster as determined from a single temperature fit;
$r_c$ is the characteristic radius of the temperature profile
(see Eq.~(\ref{eqn:tfit});
$r_{\mathrm{cool}}$ is the cooling radius for a cooling time of 15~Gyr
and $a$ is the power law index of the scaling of $r_{\mathrm{cool}}$ with
cooling time (see text for details). }
\label{tab:temps}
\centerline{
\begin{tabular}{|lccrrr|}
\hline
cluster     & $T_h$ & $(T_h-T_c)$ & $r_c$ & $r_{\mathrm{cool}}$ & $a$ \\ 
            & (keV) &   / $T_h$   & (kpc) & (kpc)               &     \\
\hline
NGC 533          & 1.29$\pm$ 0.08& 0.53$\pm$ 0.03& 21 &  50& 0.76\\
Virgo            & 2.63$\pm$ 0.09& 0.47$\pm$ 0.02& 18 &  73& 0.86\\
A 262            & 2.24$\pm$ 0.09& 0.58$\pm$ 0.02& 25 &  85& 0.89\\
A 1837           & 4.41$\pm$ 0.35& 0.41$\pm$ 0.03& 50 &  49& 1.05\\
S\'ersic 159$-$3 & 2.43$\pm$ 0.06& 0.19$\pm$ 0.00& 27 & 174& 0.66\\
MKW 9            & 2.61$\pm$ 0.40& 0.54$\pm$ 0.08& 65 &  58& 0.78\\
2A 0335+096      & 3.04$\pm$ 0.15& 0.57$\pm$ 0.03& 62 & 166& 0.66\\
MKW 3s           & 3.54$\pm$ 0.15& 0.18$\pm$ 0.01& 45 & 130& 0.79\\
A 2052           & 3.12$\pm$ 0.10& 0.72$\pm$ 0.02& 26 & 126& 0.78\\
A 4059           & 4.05$\pm$ 0.17& 0.58$\pm$ 0.02& 42 & 118& 0.75\\
Hydra A          & 3.36$\pm$ 0.12& 0.15$\pm$ 0.01& 41 & 177& 0.75\\
A 496            & 4.38$\pm$ 0.31& 0.53$\pm$ 0.04& 74 & 123& 0.72\\
A 3112           & 4.48$\pm$ 0.14& 0.41$\pm$ 0.01& 54 & 170& 0.68\\
A 1795           & 5.82$\pm$ 0.25& 0.43$\pm$ 0.02& 99 & 176& 0.67\\
A 399            & 6.23$\pm$ 0.57& 0.73$\pm$ 0.07& 54 &  65& 1.15\\
A 3266           & 8.72$\pm$ 1.13&         -     &  - &  57& 1.02\\
Perseus          & 6.53$\pm$ 0.54& 0.55$\pm$ 0.05&147 & 177& 0.67\\
Coma             & 7.48$\pm$ 0.34&        -      &  - &   -& 0.96\\
A 754            & 7.98$\pm$ 0.79& 0.38$\pm$ 0.04&  - &  79& 0.98\\
A 1835           & 7.21$\pm$ 0.77& 0.36$\pm$ 0.04&141 & 256& 0.64\\
\hline\noalign{\smallskip}
\end{tabular}
}
\end{table}

We have also scaled our temperature profiles using the following expression:
\begin{equation}\label{eqn:tfit}
T(r) = T_c + (T_h-T_c) \frac{(r/r_\mathrm{c})^{\mu}}
{1+(r/r_\mathrm{c})^{\mu}},
\end{equation} 
which is a reasonable universal fit to the temperature profile of cooling
clusters (Allen et al.  \cite{allen01}).  In addition, the choice of this
profile is motivated by the analysis of Voigt et al.  (\cite{voigt02}).  In
order to reduce the number of parameters we set $T_c$ equal to the temperature
of the central bin and use $\mu=2$, as Allen et al.  (\cite{allen01}) find
$\mu=1.9 \pm 0.4$ when fitting the scaled temperature profiles of 6 cooling
clusters. 

The scaled temperature profile are shown in Fig.~\ref{fig:trc}.  Data outside a
radius of 5$r_c$ are not shown, as for some clusters the temperature starts
decreasing again, therefore violating (\ref{eqn:tfit}).  All clusters obey the
scaling curve (\ref{eqn:tfit}) remarkably well.  The cooling radius that we find
correlates well with $T_h$ ($r_c = (4.1\pm 0.8) T_h^{1.84\pm 0.14}$, with $r_c$
in kpc and $T_h$ in keV).  Since the virial radius scales also with $T_h^{0.5}$,
the ratio of $r_c$ to the virial radius is not constant, contrary to the
findings of Allen et al.  (\cite{allen01}).  Note however that Allen et al.
have a smaller sample with a more limited temperature range (hot, distant
clusters).

We have investigated correlations of the cooling gas with other parameters.  In
Table~\ref{tab:temps} we list the average temperature $T_h$ of the hot gas
outside the cooling radius, the relative temperature decrement $(T_h-T_c)/T_h$,
as well as the cooling radius $r_{\mathrm{cool}}$, defined here as the radius
where the cooling time $t_{\mathrm{cool}}$ equals 15~Gyr.  Over a wide range in
cooling time (typically for cooling times between 5--50~Gyr), the cooling radius
is found to scale as $r_{\mathrm{cool}}\sim t_{\mathrm{cool}}^a$.  We also list
the best-fit parameter $a$ in Table~\ref{tab:temps}.

We have searched for correlations of the relative temperature decrement
$(T_h-T_c)/T_h$ with various quantities such as the cooling radius
$r_{\mathrm{cool}}$, the hot gas temperature $T_h$, the average temperature
gradient $T_h/r_{\mathrm{cool}}$, the central cooling time (hence central
density), but none of them are significant.  There is only a weak
anti-correlation between cooling radius and relative temperature decrement, in
the sense that the larger cooling regions tend to have relatively weaker cooling
flows.  However, this is only significant at the 10~\% significance level, and is
perhaps a selection effect in the sense that the larger clusters tend to have
lower densities hence longer cooling times.  Another explanation is that the
larger clusters have a relatively inefficient heat conduction (proportional to
${\mathrm d}T/{\mathrm d}r$).  Therefore it is more difficult to balance
radiative losses by heat conduction in the more extended clusters.  The role of
heat conduction is investigated in more detail in Sect.\ref{sect:heatcon}.

We note that most clusters have a rather narrow range of temperature decrement,
between 0.35--0.60.  The only exceptions are A~399 and A~2052, both with a high
temperature decrement around 0.72, and S\'ersic~159$-$3, MKW~3s and Hydra~A, all
three with a low temperature decrement of about 0.2.  These temperature
decrements are in excellent agreement with the results from the high-resolution
RGS spectra of these clusters (Peterson et al.  \cite{peterson03}).

The most important finding of our paper is however that the spectra of the
cooling region are not isothermal.  They are not isothermal in two aspects:
first, the maximum temperature at each radius decreases towards the center;
secondly, at each radius the gas has a range of temperatures, the emission
measure distribution at each radius showing a steep decline or cut-off towards
lower temperatures.

The decline of the maximum temperature of the gas towards the center of the
cluster in all our cooling clusters (Table~\ref{tab:fitpar1}--\ref{tab:fitpar2})
excludes a class of models where the core consists of a mixture of hot
isothermal gas and significantly cooler gas, with a position-dependent filling
factor of the cool gas (for example Fukazawa et al.  \cite{fukazawa94}; Ikebe et
al.  \cite{ikebe99}).

The fact that even at a single radius the spectrum is not isothermal contradicts
any model for a spherical symmetric, single phase cooling flow model.  We will
elaborate on this last model in the next section.  Here we note that the
non-isothermality is not caused by the width of our extraction annuli, because
the temperature gradients across our annuli are not large enough for that.

Most of our clusters have a small but significant emission measure down to about
0.5$T_{\max}$ (Table~\ref{tab:fitpar1}--\ref{tab:fitpar2}).  Our fits to this
differential emission measure distribution with a cut-off isobaric cooling flow
model (Sect.~\ref{sect:multit}, Table~\ref{tab:fittemp}) show a range of values
for the ratio $T_{\min}/T_{\max}$ for the cases with good statistics.  As we
show in this paper, the isobaric cooling flow model does not apply to our
clusters but nevertheless the shape of its emission measure distribution is a
convenient parameterisation of the shape of the actual emission measure
distribution.  This temperature ratio $T_{\min}/T_{\max}$ varies between
0.7--0.9 for a cluster like Hydra~A, 0.46 for Coma and A~262, down to 0.30 for
2A~0335+096 and 0.25 for Perseus.  Alternatively, the minimum temperature
$T_{\min}$ can be as low as 0.7~keV for NGC~533 and A~262 but at least twice as
large for other clusters.

There is no clear correlation between this cut-off temperature (both in absolute
and relative terms) and other cluster parameters.  This excludes for example
inhomogeneous metallicity models, as shown already by Peterson et al.
(\cite{peterson03}) from the RGS data of these clusters.

Our fits to the cut-off isobaric cooling flow model have relatively large error
bars.  This is due to the fact that our results are based upon a two-stage
fitting procedure:  first determine the differential emission measure
distribution, and from that derive the best-fit cut-off isobaric cooling flow
parameters.  In particular the temperature of the hottest component $T_1$
sometimes has a rather large uncertainty despite the good statistical quality of
the spectrum.  This is due to the fact that by increasing $T_1$ its emission
measure $Y_1$ should decrease, but because the cooler temperature components
also shift their temperatures, a rearrangement of the emission measure
distribution takes place resulting in almost the same spectrum and differential
emission measure distribution, but as a result with relatively large error bars.
This implies that our test for consistency with an isothermal model as indicated
by an "I" in Tables~\ref{tab:fitpar1}--\ref{tab:fitpar2} is not the most
sensitive test.

With respect to that our fits with the emission measure distribution
(\ref{eqn:demalpha}) is more robust.  It shows that in general the emission
measure distribution of the cooling gas has a rather constant shape, both as a
function of radius as well as from cluster to cluster (Sect.~\ref{sect:wdem}).
We come back to this in our discussion on magnetic loop models
(Sect.~\ref{sect:loopmodels}).

It is impossible to discuss in this paper all models proposed to solve the
cooling flow problem in relation to our current findings.  In the next sections
we discuss a few of the most promising scenario's.  But we hope that the
model independent emission measure distributions derived in this paper will be
useful to test any alternative cooling flow model.

\subsection{The isobaric cooling flow model}

\begin{figure}
\resizebox{\hsize}{!}{\includegraphics[angle=-90]{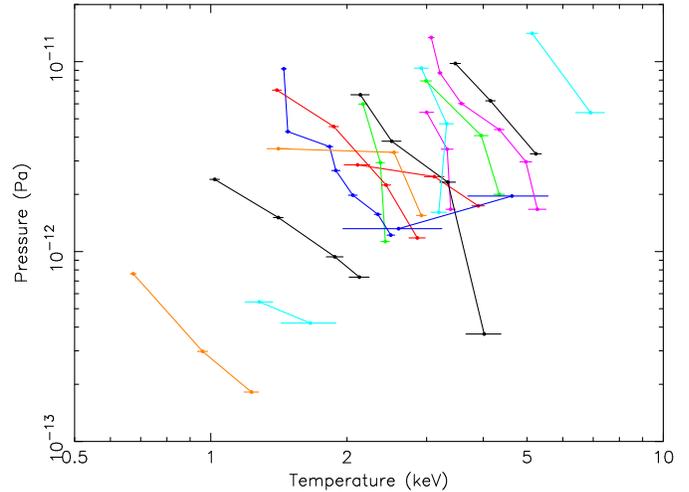}}
\caption{Pressure versus temperature for 16 clusters. Only results for shells
where the cooling time is less than 15~Gyr are shown. The center of the cluster
is the leftmost data point for each cluster. Pressure and temperature are
derived from the isothermal model fits (Table~\ref{tab:fit1t}).}
\label{fig:pt}
\end{figure}

It has been argued that a simple stationary flow dominated by radiative cooling
is inconsistent with the density profiles of clusters (Fabian et al.
\cite{fabian84}, Thomas et al.  \cite{thomas87}, Johnstone et al.
\cite{johnstone92}).  If radiative cooling were the only important energetic
process, we would expect that the emission measure distribution would resemble
that of (\ref{eqn:isocf}) at each radius.  Clearly, the high resolution RGS
observations are inconsistent with that distribution integrated over the entire
cluster (Kaastra et al.  \cite{kaastra01}; Peterson et al.  \cite{peterson01};
Tamura et al.  \cite{tamura01a}, \cite{tamura01b}; Peterson et al.  2003
\cite{peterson03}).

Detailed measurements of the temperature distribution at each radius have been
relatively difficult until observations with XMM-Newton EPIC and Chandra.  Many
recent studies (David et al.  \cite{david01}, Molendi \& Pizzolato
\cite{molendi01}, Johnstone et al.  \cite{johnstone02}, Matsushita et al.
\cite{matsushita02}, Schmidt et al.  \cite{schmidt02}, Sanders \& Fabian
\cite{sanders02}, Buote \cite{buote02}) have argued to varying degrees that the
temperature distribution at each radius is isothermal or very close to
isothermal at each radius.  Our measurements are in broad agreement with the
average temperatures found by these studies.  We argue, however, that a
single-phase temperature distribution at each radius is unlikely to be the case
as demonstrated by the better spectral fits with our multi-temperature or {\sl
wdem} models.  Instead, either a relatively steep or narrow differential
emission measure distribution might be required.  The exact form of the
differential distribution is difficult to determine at the current spectral
resolution.  It might be possible that the parameterization of the global
temperature distribution offered in Peterson et al.  (\cite{peterson03}) partly
reflects a temperature gradient as in (\ref{eqn:tfit}) but also represents a
steep intrinsic differential emission distribution like that modelled in
Sect.~\ref{sect:wdem}.

In any case the present data rule out a model with a local isobaric cooling flow
(\ref{eqn:isocf}) at each radius for most of our clusters.  The few cases where
we cannot rule it out formally are those cases where the spectra are relatively
noisy.  This is evident from the steep emission measure distribution or cut-off
emission measure distribution that we find in the present work.

Finally, our spectral modeling confirms that the cluster cores are not isobaric:
for most of our clusters, the pressure drops by half an order of magnitude from
the cluster center to the cooling radius $r_{\mathrm{cool}}$
(Fig.~\ref{fig:pt}).

\subsection{Heating by AGN}

\begin{table}[!ht]
\caption{Radio fluxes of the central galaxies in clusters. $S_{1.4}$ is
the flux (mJy) at 1.4~GHz, $L_{\nu}$ the intrinsic luminosity at the same
frequency, $\alpha$ is the radio spectral index ($L_{\nu}\sim \nu^{\alpha}$).
$L_{\mathrm X}$ is the total 0.2--10~keV luminosity within the cooling
radius $r_{\mathrm{cool}}$ as given in Table~\ref{tab:temps}.
}
\label{tab:radio}
\centerline{
\begin{tabular}{|@{\,\,}l@{\,\,}l@{\,}r@{\quad}r@{\quad}r@{\quad}r@{\,\,}|}
\hline
Cluster & Central & $\log S_{1.4}$ & $\log \nu L_{\nu}$ & $\alpha$ & 
    $\log L_{\mathrm X}$\\
        & galaxy  & (mJy)          & (W)   & & (W) \\
\hline
             NGC 533&NGC 533         &     1.5&     31.7&      & 35.3\\
               Virgo&M 87            &     5.3&     33.9&  -0.8& 36.2\\
               A 262&NGC 708         &     1.8&     32.0&  -0.8& 36.2\\
              A 1837&GSC 5557-266    &     0.7&     32.2&      & 36.0\\
    S\'ersic 159$-$3&MCG $-$7-47-23  &     2.4&     33.8&  -1.3& 37.5\\
               MKW 9&UGC 9886        &  $<$0.3&  $<$31.3&      & 35.5\\
         2A 0335+096&EQ 0335+096     &     1.6&     32.4&      & 37.7\\
              MKW 3s&NGC 5920        &     2.1&     33.2&  -2.0& 37.2\\
              A 2052&UGC 9799        &     3.7&     34.6&  -1.3& 37.1\\
              A 4059&MCG $-$6-1-8    &     3.1&     34.2&  -1.7& 37.0\\
               A 780&Hydra A         &     4.6&     35.9&  -0.8& 37.7\\
               A 496&MCG $-$2-12-39  &     2.1&     32.9&  -0.8& 37.2\\
              A 3112&ESO 248-6       &     3.2&     34.7&  -0.6& 37.8\\
              A 1795&MCG $+$5-33-5   &     3.0&     34.4&  -0.9& 37.9\\
               A 399&UGC 2438        &  $<$0.3&  $<$31.8&      & 36.1\\
              A 3266&ESO 118-30      &  $<$0.7&  $<$32.1&      & 36.0\\
     Perseus        &NGC 1275        &     4.1&     34.4&  -0.8& 38.0\\
Coma                &NGC 4874        &     2.3&     32.9&  -0.8& -\\
               A 754&LEDA 25777      &     0.9&     32.2&      & 36.3\\
              A 1835&A 1835-1        &     1.6&     34.2&      & 38.8\\
\hline\noalign{\smallskip}
\end{tabular}
}
\end{table}

It has been proposed that heating by a central AGN could balance, at least for a
part, the radiative losses in a cluster due to cooling (e.g.  B\"ohringer et al.
\cite{boehringer02}).  In order to test this hypothesis on our large sample, we
have collected the radio fluxes of the central galaxies in each cluster
(Table~\ref{tab:radio}).  The fluxes are all evaluated at 1.4~GHz (20~cm) and
most of them were taken from the NRAO VLA Sky Survey (Condon et al.
\cite{condon98}) or other surveys available in the literature.

We have searched for a correlation of the relative temperature decrement
$(T_h-T_c)/T_h$ with the radio luminosity of the central galaxy, but there is no
significant correlation (Spearman rank correlation coefficient $-$0.28, but
formally significant only at the 1.1$\sigma$ level).  Only the most radio
luminous cluster in our sample (Hydra A) is also the cluster with the smallest
temperature decrement, despite the short central cooling time of 1.6~Gyr.

\begin{figure}
\resizebox{\hsize}{!}{\includegraphics[angle=-90]{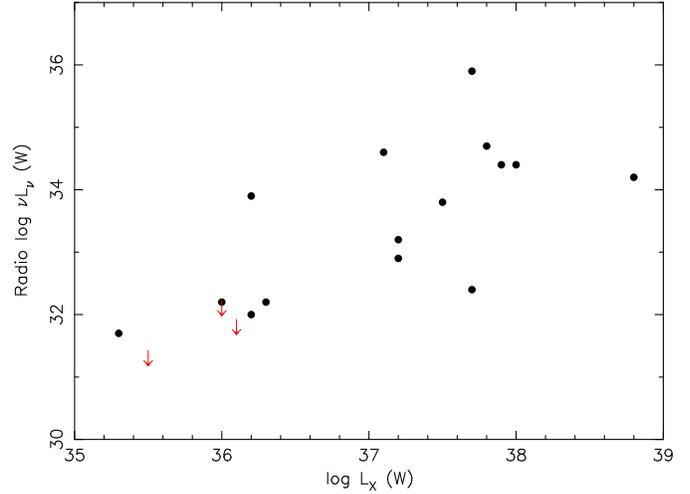}}
\caption{Correlation between the 0.2--10~keV X-ray luminosity within
the cooling radius $r_{\mathrm{cool}}$ as given in Table~\ref{tab:temps}
and $\nu L(\nu)$ at 1.4~GHz for the central radio galaxy. Upper limits
are indicated by an arrow.}
\label{fig:cfradio}
\end{figure}

However there is a correlation between the radio luminosity of the central AGN
with the X-ray luminosity within the cooling radius (Fig.~\ref{fig:cfradio}).
Interestingly, such a correlation was not found in a similar analysis based upon
Chandra data (Voigt \& Fabian \cite{voigt03}), however they used luminous
cooling clusters (X-ray luminosity within the cooling radius $>10^{37}$~W),
while we also include less luminous cooling clusters.  This correlation could be
consistent with a heating by the central AGN scenario, but a correlation alone
is insufficient proof for this.  First, the correlation may be due to some
selection effect or common "hidden" variable to which both $L_X$ and
$L_{\mathrm{AGN}}$ correlate.  Furthermore, radio flux alone is not a proper
measure of the time averaged power of an AGN.  The relation between total power
and observed radio luminosity is not perfect, due to intrinsic differences in
radio loudness and orientation (beaming) effects, as well as the fact that radio
and X-rays look at different time scales.

\subsection{Heat conduction by electrons\label{sect:heatcon}}

Thermal conduction by electrons might play an important role in cooling
clusters.  It is possible that some heat influx from the hot outer region of the
cluster to the center, totally or only partially balances radiative losses.  We
have shown in Sect.~\ref{sect:multit}, however, that the ICM is not locally
isothermal at each radius but has a range of temperatures with a steep
differential emission measure.  Although the cause of this particular emission
measure distribution is unknown, it is unlikely that thermal conduction would be
effective on large (Mpc) scales from the outer regions of clusters if conduction
is incapable in removing temperature inhomogeneities on smaller (kpc) scales.
This seems unlikely since as the conduction coefficient is increased, thermal
instability is removed first at the largest wave numbers (Field \cite{field65}).

We can also derive conduction coefficients ignoring the above argument and
assuming that heat conduction by electrons balances radiative losses, i.e.: 
\begin{equation}
\label{flux} 
\int_{\mathrm{V}}n^2 \Lambda(T)d \,
V=\int_{\mathrm{S}} \kappa (\nabla T)d \,S, 
\end{equation} where $n$, $\Lambda$,
$T$ and $\kappa$ are the gas density, cooling function, temperature and thermal
conductivity, respectively and $V$ and $S$ are the volume and surface area of
the X-ray emitting region, respectively.  Assuming spherical symmetry and using
data binned into shells, the conductivity coefficients at the boundary of each
shell can be computed.  Using this procedure, Voigt et al.  (\cite{voigt02})
investigated thermal conduction in two cooling clusters:  A\,2199 and A\,1835.
We extent this analysis to our sample, including the three non cooling clusters
as a control case.  The temperature gradient between shells is evaluated by
fitting the temperature profile with the function (\ref{eqn:tfit}).  We exclude
A\,1837 from the analysis because its temperature profile cannot be fitted with
(\ref{eqn:tfit}), since only two bins are within the cooling radius.

\begin{figure}
\resizebox{\hsize}{!}{\includegraphics{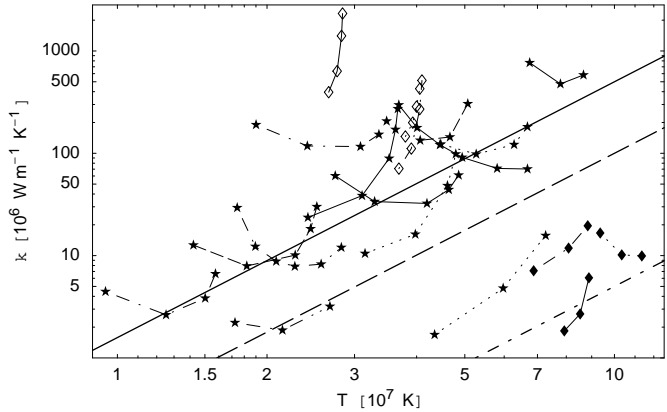}}
\caption{The conduction coefficients $\kappa$ required for heat conduction to
balance radiation losses as a function of temperature.  The solid line is the
Spitzer conductivity, the dashed lines are one fifth and one hundredth of the
Spitzer conductivity, respectively.  Filled diamonds represent the values for
the three clusters without cooling flow (Coma, A\,3266 and A\,754), while open
diamonds represent the values for the clusters with a shallow temperature
profile (MKW\,3s, S\'ersic 159-3 and Hydra\,A).  Filled stars represent the
remaining clusters.  Values for the same cluster are joined by a line and values
are given only for bins with cooling times less than $\sim$30~Gyr.}
\label{fig:kappa}
\end{figure}

Our estimates of the conduction coefficients $\kappa$ are shown in
Fig.~\ref{fig:kappa}, in which we do not plot error bars in order to facilitate
the reading (the 1$\sigma$ errors for $\kappa$ are $\sim$20--90~\% depending
mainly on the accuracy of the temperature estimates).

The estimated conduction coefficients must be compared to theoretical
calculations to see whether heat conduction from the outer regions can totally
balance radiative losses.  In this paper, we do not discuss the possibility of a
partial balance of heat conduction and radiative losses.  For a highly ionised
plasma such as the ICM, the maximum rate is expected to be the Spitzer
conductivity $\kappa_{\mathrm{s}}$ (Spitzer \cite{spitzer62}).  In the presence
of a homogeneous magnetic field, the conductivity is the Spitzer rate only along
the field, but severely decreased in the transverse direction.  Thermal
conduction in a tangled magnetic field has been studied by Chandran \& Cowley
(\cite{chandran98}).  They conclude that in cooling flows, thermal conductivity
is below the Spitzer level by a factor of order $10^2$ to $10^3$, depending on
the assumed field tangling scale.  Conductivity is less severely decreased if
the magnetic field behaves chaotically over a wide range of scales:  Narayan \&
Medvedev (\cite{narayan01}) estimate that in this case conductivity is only a
factor $\sim$5 below the Spitzer rate (given the uncertainties this factor lies
in the range $\sim$2.5--10).  Therefore, the Spitzer conductivity
$\kappa_{\mathrm{s}}$, $\kappa_{\mathrm{s}}/5$ and $\kappa_{\mathrm{s}}/100$ are
shown in Fig.~\ref{fig:kappa} for comparison.

For the three non cooling clusters we find very low values for the conductivity
coefficients:  the absence of significant cooling allows even inefficient heat
conduction to remove temperature inhomogeneities.  For the cooling clusters we
see that clusters with shallow temperature profile show a very different trend
and relatively high conductivity because of the small temperature gradients.
In addition, in most of the clusters conductivity increases towards the center
(lowest temperature) and as we approach the radius at which the temperature
reaches its maximum (this can not be clearly seen in Fig.~\ref{fig:kappa} for
every cluster because for some the cooling time at this radius is larger than
$\sim$30~Gyr and consequently not shown).  This trend is due to the fact that
our temperature profile model has vanishing gradient both at large and small
radii.  Most important is the fact that we find conductivities larger that one
fifth of the Spitzer rate in almost all the cooling clusters.  Since the thermal
conductivity in the ICM is estimated to be below the Spitzer rate by at least a
factor of 5, we conclude that heat conduction alone is insufficient to balance
radiative losses in cooling clusters.

Although we demonstrated that heat conduction on a global scale cannot maintain
the large scale temperature gradients in cooling clusters, heat conduction 
may still play an important role on smaller spatial scales.  In the next section
we consider a class of magnetic loop models as a possible mechanism in which 
heat conduction on smaller scales may occur.

\subsection{Magnetic loop models\label{sect:loopmodels}}

\subsubsection{Cluster magnetic fields}

It has been argued in the past by several groups that magnetic fields may play
an important role in the dynamics of clusters.  For example, Clarke et al.
(\cite{clarke01}) show in a study of rotation measures for a large sample of
X-ray bright Abell clusters that the magnetic field occupies at least 95~\% of
the area of the non-cooling innermost 750~kpc of these clusters.  This suggests
that enhanced magnetic field levels permeate the clusters with a high filling
factor, with magnetic structure down to the level of 10~kpc or even smaller.
The average magnetic field is of the order of 1~nT, and the corresponding
magnetic energy density is a few percent of the thermal energy density of the
gas.  In the centers of cooling clusters, magnetic fields up to 4~nT are found
(Taylor et al.  \cite{taylor02}).

One of the most important aspects of magnetic fields in clusters of galaxies is
that heat conduction in order to balance radiative losses can only occur along
the magnetic field lines.  Thus, the effectiveness of thermal conduction depends
strongly upon the magnetic topology.  For example, Norman \& Meiksin
(\cite{norman96}) have considered models where relatively cold loops reconnect
with hot loops that are connected to the thermal reservoir of the cluster.  This
process would effectively slow down the cooling.  A problem with this model is
that hot loops must extend down from the outer parts of the cluster to the
region where the cool loops reside.  In that case, we would still expect --
apart from the cooling gas -- to see a hot component throughout the cluster
volume with an almost constant temperature.  In all cooling clusters we find
however that the temperature of the hottest gas component decreases
significantly towards the center of the cluster.

\subsubsection{Coronal loop models}

Here we consider a different class of magnetic loops, motivated by a recent
paper by Aschwanden \& Schrijver (\cite{aschwanden02}, AS hereafter).  These
authors created an analytical model for solar and stellar coronal loops.  The
differential emission measure distribution of their models resembles the
differential emission measure distributions that we obtained for our clusters:
a distribution that is strongly peaked towards the maximum temperature $T_{a}$
at the top of the loops.

Their model considers stationary loops (i.e.  no net gas flows) for which mass
and momentum conservation is taken into account.  In the energy balance
equation, radiative losses, heat conduction and position-dependent heating are
taken into account, as well as the effects of gravity.  For the geometry of the
loops they took a semicircular shape with height $h(s)=h_1\sin(s/h_1)$ where $s$
is the coordinate along the loop.  The loop half length is $L$ and the loop
crosses the photosphere at $s=s_0$.  The loop cross section $a(s)$ expands by an
expansion factor $\Gamma$ from the photosphere to the loop apex.  The heating
process is not prescribed in the AS models, it is only parameterised with a form
$E_H(s) = E_0 \exp{(-s/s_H)}$, with $E_0$ the heating rate (W\,m$^{-3}$) at the
foot points of the loop and $s_H$ a scale height.  For $s_H=0$, all heating
takes place at the loop foot points, while for $s_H \gg L$ the heating occurs
uniformly across the loop.  The hydrodynamic equations are solved with as
boundary condition the requirement that the conductive flux at the foot points
and top of the loop vanishes.  The fact that ${\mathrm d}T/{\mathrm d}s=0$ at
the apex of the loops causes the sharp peak of the differential emission measure
just below the apex temperature $T_{a}$.  In general, the solutions have a
temperature decrease and density increase from apex to foot points, with only a
relatively small pressure increase.  The loops also obey scaling laws, which are
to lowest order similar to the scaling laws derived first by Rosner et al.
(\cite{rosner78}), namely $n_0 L\sim T_{a}^2$, with $n_0$ the gas density at the
loop apex.  At fixed $T_{a}$, the loops have therefore similar column density
($\sim n_0 L$), but the shorter loops have higher density and therefore are
relatively very efficient X-ray emitters (emissivity scales with $n_0^2
\Lambda(T)$).

\subsubsection{Application of the coronal loop models to clusters}

We expect the same physical processes that play a role in coronal loops to play
a role in magnetic loops in clusters, be it on a different scale.  The centers
of clusters contain highly tangled magnetic fields, hence heat conduction takes
preferentially place along the magnetic field lines of the loop.  Also the
densities are high enough that radiative cooling occurs.  Loops can be heated by
a variety of processes, whatever their origin.  In fact the lack of large
amounts of cool gas in the cluster cores provides evidence for (effective)
heating, either direct or indirect by mixing processes.  The precise heating
process is not important for our present discussion.  Whatever the heating
process in clusters is, the heat will predominantly be transported along the
magnetic field lines and the heat will finally be lost due to the observed
radiative cooling.  These are exactly the same conditions that are the basis for
the solar coronal loop models.

It is harder to identify the foot points of magnetic loops in clusters.  These
foot points can be associated with active galactic nuclei, member galaxies, or
just enhancements in the highly tangled intracluster field.

\begin{table*}[!ht]
\caption{Emission measure distributions for the Aschwanden \& Schrijver
(\cite{aschwanden02}) models.  $T_a$ (keV) is the maximum temperature at the
loop apex.  $\Gamma$ is the loop cross section ratio between the loop apex and
the photosphere.  $L$ is the loop half length (kpc), and $s_H$ the heating scale
height.  $M_{13}$ is the cluster mass in $10^{13}$~M$_{\sun}$ within the radius
$R$ (kpc) where the loop is located.  $Y_1$--$Y_5$ are the emission measures
$Y_i$ (in m$^{-3}$ of the temperature components 1--5.  The corresponding
temperatures are $T_i = T_1 / 2^{i-1}$.  The density at the loop apex is $n_a$
(given in units of $10^4$~m$^{-3}$) and the average density over the loop volume
is $\overline{n}$.  The mass of an individual loop is $M_l$ (in
$10^{10}$~M$_{\sun}$).  Finally, $N$ is the maximum number of loops per scale
height, defined by $N\equiv 4\pi R^3/V$ with $V$ the volume of an individual
loop.  All calculations are done for loops with a cross-sectional radius at the
foot points ($r_0$) of 10~kpc.  For other values of $r_0$, the following scaling
applies:  $Y_i\sim r_0^2$, $M_l\sim r_0^2$ and $N\sim r_0^{-2}$.
}
\label{tab:armodel}
\centerline{
\begin{tabular}{|rrrrrr|rrrr|rrrrrr|}
\hline
$T_a$ & $\Gamma$ & $L$ & $s_H/L$ & $M_{13}$ & $R$ &
$Y_2/Y_1$& $Y_3/Y_1$& $Y_4/Y_1$& $Y_5/Y_1$ &
$Y_1$ & $n_a$ & $\overline{n}/n_a$ & $T_1/T_a$ & $M_l$ & $N$ \\
\hline                                                                              	    
 0.5& 10&  30& 10  & 4  & 200& 0.440& 0.217& 0.103& 0.049& 3.1E67& 0.01    & 1.58&0.64& 0.03  & 1720    \\
 1  & 10&  30& 10  & 4  & 200& 0.146& 0.054& 0.022& 0.009& 1.6E69& 0.07    & 1.29&0.88& 0.19  & 1720    \\
 2  & 10&  30& 10  & 4  & 200& 0.096& 0.027& 0.009& 0.003& 3.2E70& 0.38    & 1.16&0.91& 0.90  & 1720    \\
 4  & 10&  30& 10  & 4  & 200& 0.069& 0.015& 0.004& 0.001& 5.3E71& 1.63    & 1.10&0.93& 3.64  & 1720    \\
 8  & 10&  30& 10  & 4  & 200& 0.057& 0.010& 0.002& 0.001& 6.8E72& 5.93    & 1.08&0.93& 13.0  & 1720    \\
\hline
 2  &  1&  30& 10  & 4  & 200& 0.319& 0.113& 0.040& 0.014& 5.3E69& 0.40    & 1.35&0.74& 0.18  & 10700   \\
 2  &  3&  30& 10  & 4  & 200& 0.163& 0.054& 0.019& 0.007& 1.4E70& 0.39    & 1.23&0.87& 0.36  & 4770    \\
 2  & 10&  30& 10  & 4  & 200& 0.096& 0.027& 0.009& 0.003& 3.2E70& 0.38    & 1.16&0.91& 0.90  & 1720    \\
 2  & 30&  30& 10  & 4  & 200& 0.055& 0.011& 0.004& 0.001& 8.4E70& 0.38    & 1.13&0.93& 2.36  & 630     \\
 2  &100&  30& 10  & 4  & 200& 0.029& 0.004& 0.001& 0.000& 2.6E71& 0.38    & 1.11&0.94& 7.38  & 200     \\
\hline
 2  & 10&   3& 10  & 4  & 200& 0.074& 0.019& 0.006& 0.002& 5.1E71& 5.06    & 1.10&0.92& 1.13  & 17200   \\
 2  & 10&  10& 10  & 4  & 200& 0.080& 0.021& 0.007& 0.002& 1.3E71& 1.41    & 1.12&0.92& 1.06  & 5160    \\
 2  & 10&  30& 10  & 4  & 200& 0.096& 0.027& 0.009& 0.003& 3.2E70& 0.38    & 1.16&0.91& 0.90  & 1720    \\
 2  & 10& 100& 10  & 4  & 200& 0.153& 0.051& 0.018& 0.006& 4.0E69& 0.06    & 1.29&0.88& 0.56  & 516     \\
 2  & 10& 300& 10  & 4  & 200& 0.528& 0.216& 0.077& 0.027& 1.6E68& 0.01    & 1.44&0.65& 0.24  & 172     \\
\hline
 2  & 10&  30&  0.1& 4  & 200& 0.006& 0.000& 0.000& 0.000& 1.8E71& 0.95    & 1.08&1.00& 2.07  & 1720    \\
 2  & 10&  30&  0.3& 4  & 200& 0.030& 0.005& 0.001& 0.000& 6.0E70& 0.54    & 1.10&0.98& 1.20  & 1720    \\
 2  & 10&  30&  1  & 4  & 200& 0.060& 0.013& 0.004& 0.001& 4.0E70& 0.43    & 1.13&0.95& 0.99  & 1720    \\
 2  & 10&  30&  3  & 4  & 200& 0.090& 0.024& 0.008& 0.003& 3.3E70& 0.38    & 1.16&0.92& 0.90  & 1720    \\
 2  & 10&  30& 10  & 4  & 200& 0.096& 0.027& 0.009& 0.003& 3.2E70& 0.38    & 1.16&0.91& 0.90  & 1720    \\
\hline
 2  &  6&   2& 10  & 4  & 200& 0.096& 0.027& 0.009& 0.003& 5.0E71& 7.65    & 1.12&0.91& 0.74  & 40300   \\
 2  & 10&  30& 10  & 4  & 200& 0.096& 0.027& 0.009& 0.003& 3.2E70& 0.38    & 1.16&0.91& 0.90  & 1720    \\
 2  & 30& 102& 10  & 4  & 200& 0.096& 0.026& 0.009& 0.003& 8.7E69& 0.06    & 1.23&0.91& 1.38  & 185     \\
\hline
 2  & 10&  30& 10  & 4  & 200& 0.096& 0.027& 0.009& 0.003& 3.2E69& 0.38    & 1.16&0.91& 0.90  & 1720    \\
 2  & 10&  30& 10  & 0  & 200& 0.072& 0.018& 0.006& 0.002& 5.4E70& 0.52    & 1.09&0.92& 1.16  & 1720    \\
 2  & 10&  30& 10  & 0.8& 100& 0.089& 0.024& 0.008& 0.003& 3.7E70& 0.41    & 1.14&0.91& 0.96  & 215     \\
\hline\noalign{\smallskip}
\end{tabular}
}
\end{table*}

We used the model of AS and applied it for typical cluster conditions.  Typical
median masses of $8\times 10^{12}$~M$_{\sun}$ at 100~kpc and $4\times
10^{13}$~M$_{\sun}$ at 200~kpc were used (derived from the work of Ettori et al.
\cite{ettori02}).  We took the differential emission measure distribution from
the AS models, and binned it into bins with a factor of two difference in
temperature, just as we did for our clusters.  The highest temperature bin was
centered around the emission measure weighted loop temperature, which is close
below the maximum temperature $T_a$ at the apex of the loop.
Table~\ref{tab:armodel} lists a few typical cases of the expected emission
measure distribution.

The table shows that the shape of the emission measure distribution (as
characterised for example by the ratio $Y_2/Y_1$) depends strongly upon the
maximum temperature $T_a$ as well as the loop expansion factor $\Gamma$:  steep
differential emission measure distributions are obtained for large $T_a$ or
large $\Gamma$.  The loop length $L$ is not important for the shape of the
emission measure distribution as long as the loops are smaller than the scale
height for the gravity of the cluster (set by $R$).  The heating scale height
$s_H$ is only important when $s_H$ becomes smaller than the loop length $L$,
i.e.  when the heating occurs preferentially near the foot points of the loop
instead of uniformly over the volume.  The results do not depend strongly upon
the central mass $M_{13}$ and the position in the cluster $R$ (i.e.  the gravity
of the cluster is relatively unimportant, at least for sufficiently small
loops).  It is also seen that for a fixed temperature $T_a$, different
combinations of $\Gamma$ and $L$ can produce the same emission measure
distribution.

An important difference between various models yielding the same emission
measure distribution shape is the absolute value of the emission measure as well
as the total mass associated with the loops.  For some models, the total mass of
the loops assuming unity filling factor ($NM_l$) exceeds the gravitating mass
within that radius.  This occurs when $L$ or $\Gamma$ becomes too small.  These
unphysical models are to be rejected, hence in practice useful lower limits for
$\Gamma$ or $L$ can be obtained.

The ratio $Y_2/Y_1$ for the inner parts of our cluster sample (defined here as
the part where the cooling time is less than 10 Gyr) has a median value of 0.11,
with individual cases typically between 0.00 and 0.40.  Virgo, which has one of
the best measured spectra, has $Y_2/Y_1$ distributed between 0.05--0.13.

For large heating scale heights $s_H/L=10$ (almost uniform heating as a function
of loop radius), this ratio can be obtained for loop expansion factors $\Gamma$
of the order 10, with a typical spread of half a decade.  In this scenario, an
increase of $Y_2/Y_1$ towards the center of a cluster, where the ambient
temperature is lower, is expected if all other parameters (including $\Gamma$)
remain constant.  This is indeed observed in some of our clusters (NGC~533,
A~262 and A~2052, see Sect.~\ref{sect:wdem}).  If a uniform value of $\Gamma$
for each cluster would apply, then we would expect that the hotter clusters have
smaller $Y_2/Y_1$, i.e.  are closer to isothermal.  This is not consistent with
the observations, which show a tendency for the hotter clusters to have smaller
power law indices $\alpha$ (see Fig.~\ref{fig:overalfa}).  This then would imply
that the hotter clusters should have either smaller loop expansion factors
$\Gamma$, maybe even down to $\Gamma=1$, or larger loop length $L$.  A larger
loop length seems more likely since this keeps the total mass of the loops
within acceptable limits.

Another possibility is that the heating scale height $s_H$ is substantially
smaller than the semi loop length $L$.  In that case, the observed values for
$Y_2/Y_1$ can be explained by a $\Gamma$ that is a few times smaller than in the
case of uniform heating.  The increase of $Y_2/Y_1$ for hotter clusters could
also be explained if the hotter clusters have loops that are more uniformly
heated.

\begin{table}[!ht]
\caption{Application of the coronal loop models of Aschwanden \& Schrijver
(\cite{aschwanden02}) to the innermost three shells of 2A~0335+096. 
$M_{13}$ is the total cluster mass in $10^{13}$~M$_{\sun}$ within the radius
$R$ (kpc), for which we take the average radius of the shell.
$\overline{\rho_l}/\rho_{\mathrm{tot}}$ is the average mass density of the loops
divided by the total mass density of the cluster (including dark matter etc.).
The total mass density profile was taken from Ettori et al. (\cite{ettori02}).
In this table, $f$ is the volume filling factor of the loops.
}
\label{tab:2a0335loop}
\centerline{
\begin{tabular}{|l|rrr|}
\hline
 shell    & 1 & 2 & 3 \\
$M_{13}$  & 0.0025 & 0.07 & 0.5 \\
$R$ (kpc) & 14 & 43 & 86 \\
$\Gamma$  & 4--6 & 7--40 & 1--100 \\
$L$ (kpc) & 3.6$f^{0.5}$ & 11$f^{0.5}$ & 44$f^{0.5}$ \\
$\overline{\rho_l}/\rho_{\mathrm{tot}}$ & 0.64$f^{0.5}$ & 0.33$f^{0.5}$ & 0.15$f^{0.5}$\\
$N$ & 7--11\,$f^{0.5}$ & 14--70\,$f^{0.5}$ & 12--500\,$f^{0.5}$ \\
\hline\noalign{\smallskip}
\end{tabular}
}
\end{table}

The number of loops can be quite large.  As an example we have worked out here
the innermost three shells of 2A~0335+096.  The cooling time in these shells is
less than the age of the universe, and all three shells have significant
multi-temperature structure.  From the emission measure distribution of
Table~\ref{tab:fitpar1} we obtain the loop parameters as listed in
Table~\ref{tab:2a0335loop}.  For unity volume filling factor $f$, the loops have
a typical length of 0.25--0.50 times the radius $R$ where the loops are located.
In the outermost shell this is in fair agreement with the magnetic field
reversal length as obtained from numerical simulations of non-cooling regions
(e.g., Dolag et al.  \cite{dolag02}), in the innermost shell it is significantly
smaller.  Since the mass density of the loops compared to the total gravitating
mass density ($\overline{\rho_l}/\rho_{\mathrm{tot}}$) is relatively large for
$f=1$, a smaller volume filling factor is more likely in the innermost two
shells, giving even a much smaller loop length $L$.  A smaller filling factor of
the hot gas in the cores of several clusters is also indicated by observations
of cavities in clusters with the Rosat HRI or Chandra (for example Perseus,
B\"ohringer et al.  \cite{boehringer93}, Fabian et al.  \cite{fabian00}).
However, the number of loops needed is large, considering also that for smaller
loop length the loop radius will be also smaller than our adopted 10~kpc.  The
large number of loops as well as their small size then imply a highly tangled
magnetic field structure in the central part of the cluster.  A large number of
loops also causes a relatively smooth surface brightness, consistent with
detailed X-ray mapping of several clusters.

Some caution is needed, however.  In the context of the loop model, it is not
clear how much emission is due to the magnetised loops and how much is due to
emission from gas in between.  If a substantial fraction of the cluster emission
is from very hot gas in between the loops (for example in cavities), then in the
measured ratio $Y_2/Y_1$ the emission measure $Y_1$ should first be corrected
for this hot gas.  However, in that case there would still be the problem why we
see so little cooling gas in our clusters.

Other points of caution are the different abundances in the cluster as compared
to the solar corona, and hence a somewhat different cooling function,
and the precise physical conditions at the foot points of the loops.

\section{Conclusions}

We have determined in our work the spatially resolved differential emission
measure distribution of sample of 17 cooling clusters, plus three non-cooling
clusters for comparison.

Most of the cooling clusters show a central temperature decrement of a factor of
$\sim$2.  A few clusters (S\'ersic~159$-$3, MKW~3s and Hydra~A) do not show such
a strong temperature decrement.  In the last case this may be caused by the
luminous AGN at the center of the cluster.  Two clusters have a very strong
temperature decrement (A~399 and A~2052), the reason for this is not clear.
We also find a correlation between the X-ray luminosity within the cooling
radius with the radio luminosity of the central AGN.

It is clear from our modeling that the isobaric cooling flow model does not work
in all our clusters in regions with cooling time less than about 15~Gyr.  It
does not work on a global scale, since there is a weak pressure gradient in the
core and the spectra at each radius are not isothermal.  It also does not work
on a local scale, because there is a steep power law like emission measure
distribution at each radius.  This is much steeper than is expected from the
isobaric cooling flow model.  At each radius the differential emission measure
distribution is approximately proportional to $T^{3.1\pm0.2}$.

This emission measure distribution looks remarkably similar to the predicted
emission measure distribution of coronal loops.  Applying coronal loop models to
our emission measure distribution shows that moderate loop expansion factors of
order 10 are required (i.e., the loop tops are about a factor of 3 larger than
the foot points).  Assuming a base loop radius of order 10~kpc (the typical size
of a galaxy, or the characteristic size of the magnetic field as derived from
Faraday rotation measurements), at least a few hundred loops per shell should be
present.

\begin{acknowledgements}
This work is based on observations obtained with XMM-Newton, an ESA science
mission with instruments and contributions directly funded by ESA Member States
and the USA (NASA).  The Space Research Organisation of the Netherlands (SRON)
is supported financially by NWO, the Netherlands Organisation for Scientific
Research.  MSSL is supported financially by the UK Particle Physics and
Astronomy Research Council.  We thank an anonymous referee for making
valuable comments on the manuscript.
\end{acknowledgements}

\appendix
\section{Determination of spectra and response matrices}

We follow the usual method of accumulating spectra in concentric annuli around
the cluster center.  The accumulation of background-subtracted spectra is easy.
Background subtraction is done using the same detector regions for source and
background.  It is more difficult to obtain the proper response matrices.  They
are created as follows.  We start with a map of the intensity distribution
$\mu(x,y)$ of the cluster on the sky.  This intensity map is then multiplied by
the effective vignetting $v(x,y)$ of the telescope and detector (the off-axis
relative effective area).  This vignetted map is then convolved with the
instrumental point-spread function (psf) that can be expressed as
$p(x^\prime-x,y^\prime-y;x,y)$ giving the probability that a photon emitted at
coordinates $(x,y)$ is observed at position $(x^\prime,y^\prime)$.  In our
numerical estimates, we have taken this position dependence of the psf into
account.  The resulting map is then multiplied by the exposure map
$e(x^\prime,y^\prime)$, which gives the fraction of time that each pixel was
"on" (excluded hot pixels, gaps between CCD's and point sources are accounted
for this way).  The resulting predicted intensity distribution
$f(x^\prime,y^\prime)$ is then given by
\begin{eqnarray}
\label{eqn:convolve}
f(x^\prime,y^\prime) &=& e(x^\prime,y^\prime) \int {\mathrm d}x
                       \int {\mathrm d}y
                       \,\mu(x,y) v(x,y) \nonumber\\
                   &&    p(x^\prime-x,y^\prime-y;x,y).
\end{eqnarray}
The above formalism holds for any energy bin $E_k$.
We  divide the cluster into a set of annuli with outer radii
$r_i$. The emitted flux $\phi_i$ in annulus $i$ is then given by
\begin{equation}
\label{eqn:phi}
\phi_i \equiv 
\int\limits_{\displaystyle{0}}^{\displaystyle{2\pi}}  {\mathrm d}\theta
\int\limits_{\displaystyle{r_{i-1}}}^{\displaystyle{r_i}} r {\mathrm d}r\,
 \mu(x,y),
\end{equation}
while the observed flux $c_i$ (counts) is given by
\begin{equation}
\label{eqn:c}
c_i \equiv  
\int\limits_{\displaystyle{0}}^{\displaystyle{2\pi}}  {\mathrm d}\theta
\int\limits_{\displaystyle{r_{i-1}}}^{\displaystyle{r_i}}  r{\mathrm d}r\, 
 f(x,y).
\end{equation}
Combining (\ref{eqn:c}) with (\ref{eqn:phi}) and (\ref{eqn:convolve}),
we formally write
\begin{equation}
\label{eqn:resp}
c_i \equiv \sum\limits_{j}^{} R_{ij} \phi_j
\end{equation}
as the expected count rate in annulus $i$ due to emission in all the
annuli $j$. Again, (\ref{eqn:resp}) can be evaluated for any photon energy
$E_k$. Because the matrix $R_{ij}$ is based upon a convolution of the
source image $\mu$ with the instrumental psf $p$, the count spectrum
in each annulus depends upon the emitted spectrum in all other annuli
(i.e., the matrix $R_{ij}$ is non-diagonal).

Our initial spectral fitting method therefore consisted of making model spectra
in each annulus, and fitting all count spectra, coupled through
(\ref{eqn:resp}), simultaneously.  We also used a variant where we considered a
separate spectral model in each of a set of concentric, three-dimensional 
shells, projecting those on the sky and then convolving with the psf.

As an illustration we present in Table~\ref{tab:annucor} the predicted
correlations between the annuli for the MOS1 data of A~1795.  The magnitude of
the redistribution due to the instrumental psf as well as the vignetting effects
are clearly visible.

\begin{table}[!ht]
\caption{Distribution of the observed flux,
expressed as a percentage of the original flux emitted in a given annulus.
Calculation applies to the MOS1 data of A~1795.
The observed flux is corrected for vignetting, psf and exposure map.
The column $o$ labels the observed annulus number, the row $e$
the emitting annulus.
}
\label{tab:annucor}
\centerline{
\begin{tabular}{|r|rrrrrrrrr|}
\hline
o/e & 1 & 2 & 3 & 4 & 5 & 6 & 7 & 8 & 9\\
\hline
1& 74 &  9 &  0 &  0 &  0 &  0 &  0 &  0 &  0\\
2& 20 & 74 &  7 &  0 &  0 &  0 &  0 &  0 &  0\\
3&  3 & 14 & 82 &  7 &  0 &  0 &  0 &  0 &  0\\
4&  1 &  1 &  6 & 78 &  6 &  0 &  0 &  0 &  0\\
5&  1 &  1 &  1 &  6 & 74 &  3 &  0 &  0 &  0\\
6&  1 &  1 &  1 &  1 &  6 & 57 &  2 &  0 &  0\\
7&  0 &  0 &  0 &  0 &  1 &  2 & 44 &  2 &  0\\
8&  0 &  0 &  0 &  0 &  0 &  0 &  1 & 43 &  1\\
9&  0 &  0 &  0 &  0 &  0 &  0 &  0 &  2 & 24\\
\hline
$\Sigma$& 100 & 100 & 97 & 92 & 87 & 62 & 47 & 47 & 25\\
\hline\noalign{\smallskip}
\end{tabular}
}
\end{table}

Unfortunately this method appeared not to be very stable.  The reason is the
significant overlap of neighboring shells.  If, due to statistical fluctuations,
the best-fit normalisation of the spectrum in one shell happens to be large,
this can be compensated by decreasing the normalisation of the neighbouring
shells.  These then affect their neighbours, etc.  Apart from this effect, the
fitting procedure and in particular the error search consumes considerable
computing time, because the complex spectra of all shells need to be fitted
simultaneously, leading quickly to 50--100 free parameters of the fit, depending
on the sophistication of the spectral model.  And due to the strong correlations
between the parameters, error search takes a very long time.

Therefore we have developed a different approach.  We deproject spectra directly
and take account of the correction factors for the psf as discussed in
Sect.~\ref{sect:radprof}. We outline our method in section~\ref{sect:depro}.

\end{document}